\documentclass[12pt]{iopart}
\usepackage{graphicx}
\usepackage{bbm}

\begin{document}

\title{Revisiting the damped quantum harmonic oscillator}

\author{Stephen M. Barnett}

\address{School of Physics and Astronomy, University of Glasgow, Glasgow G12 8QQ, UK}

\ead{stephen.barnett@glasgow.ac.uk} 

\author{James D. Cresser}

\address{School of Physics and Astronomy, University of Glasgow, Glasgow G12 8QQ, UK}

\address{Department of Physics and Astronomy, University of Exeter EX4 4QL, UK}

\address{Department of Physics and Astronomy, Faculty of Science and Engineering, Macquarie University,
NSW 2109, Australia}

\ead{james.cresser@glasgow.ac.uk}

\author{Sarah Croke}

\address{School of Physics and Astronomy, University of Glasgow, Glasgow G12 8QQ, UK}

\ead{sarah.croke@glasgow.ac.uk} 

\date{\today}

\begin{abstract}
We reanalyse the quantum damped harmonic oscillator, introducing three less than common features.  These are (i) the use of a continuum model of the reservoir rather than an ensemble of discrete oscillators, (ii) an exact diagonalisation of the Hamiltonian by adapting a technique pioneered by Fano, and (iii) the use of the thermofield technique for describing a finite temperature reservoir.  We recover in this way a number of well-known and some, perhaps, less familiar results.  An example of the latter is an \emph{ab initio} proof that the oscillator relaxes to the mean-force Gibbs state.  We find that special care is necessary when comparing the damped oscillator with its undamped counterpart as the former has two distinct natural frequencies, one associated with short time evolution and the other with longer times.
\end{abstract}

\maketitle

\section*{Preface: an apology}
There can surely be no more intensively studied open quantum system than the damped harmonic oscillator. This makes it all but impossible to do justice to the vast literature on the subject, and authors seeking to write on this system need to tread carefully and to acknowledge, freely, that much of the relevant literature will be, unintentionally but inevitably, overlooked.  The task is made yet more hazardous by the fact that different communities in physics have covered similar ground more or less independently.  Nevertheless, this special issue offers the opportunity to reexamine this well-studied system, with the aim of demonstrating some techniques that should, perhaps, be better known.

\section{Introduction}
Recent technological advances make it possible to realise simple mechanical devices in the microscopic and nanoscopic regimes, the properties of which are determined by quantum effects \cite{Chan2011}.  The existence of these represents a remarkable opportunity for fundamental studies of light-matter interactions \cite{Gigan2006} and also the potential for practical application to quantum communications and information processing \cite{Stannigel2010}.  Yet they present also a challenge to existing methods of analysis, many of which were developed to treat more rapidly oscillating systems with weaker couplings.  The strong coupling regime brings with it some surprises, such as the possibility that quantum entanglement might persist in the high-temperature limit \cite{Galve2010}.

The quantum theory of machines is built, to a large extent on the theory of oscillators and strongly coupled oscillators, which are coupled to one or more environments, each of which is at a characteristic temperature \cite{Zhang2014,Joshi,Brunelli2015}.  The behaviour of these quantum systems is governed not simply by average properties but also fluctuations, and has been informed by the development of fluctuation theorems for quantum open systems \cite{Campisi,Talkner}.  Such considerations underpin developments in the rapidly advancing field of quantum thermodynamics \cite{Gemmer,Vinjanampathy2016}.

The coupling to the environment needs to be handled with some care because of the possibility of quantum coherence and the development of entanglement between the system of interest and its environment, with the result that apparently unphysical behaviours may emerge \cite{Allahverdyan}.  The requirement to treat the coupling to the reservoirs with care provides the incentive to return to the problem of a single strongly-damped oscillator and to treat this model exactly, using the techniques identified in the abstract.



There exist at present very many mathematically and physically acceptable methods for treating damped harmonic oscillators.  Common to most, if not all, of these developments is the treatment of the surrounding environment, or reservoir, as an ensemble of harmonic oscillators with a broad spectrum of oscillator frequencies.  It is the dephasing brought about by this spread of frequencies that introduces the damping, or irreversibility, in the oscillator dynamics. The environmental harmonic oscillators may be physical oscillators or vibration modes, as in the Caldeira-Leggett model \cite{CalderiaI,CaldeiraII}, or the modes of the quantised electromagnetic field, as in many quantum optics applications \cite{Senitzky,Louisell1964,Radmore1997,Louisell1973,Grabert1982,Perina,Meystre1991,Carmichael1993,Walls1994,Carmichael1999,Breuer2002,Gardiner2004,Ficek2004,Wiseman2010,Weiss2012}.

For weakly damped oscillators, such as those encountered regularly in quantum optics, there are master equations and the corresponding Heisenberg-picture operator Langevin equations.  Even in this weakly damped regime, the dissipative dynamics can be challenging with a rich structure of asymptotic states \cite{Igor1,Igor2}.  For more strongly damped systems memory effects become important and there is a departure from Markovian evolution \cite{Stenholm2001,Maniscalco2007,Haikka2011,Breuer2012,Pernice2012,Vacchini2012,Chrascinski2014,Hall2014}.  Yet stronger coupling requires the inclusion of counter-rotating interactions, which do not conserve the number of quanta. Among the many and varied approaches adopted are the aforementioned Schr\"{o}dinger-picture master equations for the oscillator density operator \cite{Hu,Diosi1993} and Heisenberg-Langevin operator equations for oscillator observables driven by environmental fluctuations \cite{Grabert1984,Ford1985,Ford1985a}.  Also widely used are Feynman-Vernon path integral and related techniques \cite{CaldeiraII,Feynman,Zinn-Justin,Kleinert,Smith1987,Grabert1988}.   For finite temperatures the environmental oscillators are considered to be prepared initially in thermal states with Bose-Einstein statistics appropriate to the reservoir temperature.  This leads, in the Schr\"{o}dinger picture, to the reservoir acting both as a source as well as a sink of quanta.  The techniques for treating this include a product of thermal density operators for the reservoir modes \cite{Louisell1964,Louisell1973,Perina,Meystre1991,Carmichael1993}, imaginary time methods and thermal Green functions \cite{Abrikosov,Lifshitz,Le Bellac} and also thermofield dynamics \cite{Takahashi,Umezawa1982,Knight1985,Dalton1987,Umezawa1993,Dalton2015}.  In addition to these methods, there has also been work done on diagonalising the oscillator-reservoir Hamiltonian in which the reservoir is formed from a collection of harmonic oscillators \cite{Haake,O'Connell}.  Our analysis develops and expands upon material in an earlier preprint \cite{preprint} (see also \cite{Huttner1992a,Huttner1992b,Dutra,PhilbinAnders}), which treats the environment as a continuum.  It is complementary to that adopted by Philbin who has tackled this problem of the oscillator evolution using Green functions \cite{Philbin2012}.  We hope that the combination of his work and ours will provide a more complete understanding.


\section{Background}

The harmonic oscillator has a special place in physics as one of the simplest and most widely employed of physical models. The reasons for its ubiquity, no doubt, are its simplicity and the fact that it is readily analysed. In the quantum domain, the harmonic oscillator is barely more difficult to treat than its classical counterpart and was one of the first dynamical systems to which Schr\"{o}dinger applied his equation \cite{Schrodinger}.  Today, both the classical and quantum forms appear in elementary courses on classical and quantum mechanics.

The damped harmonic oscillator loses energy as a result of coupling to the surrounding environment.  In the classical domain it often suffices to describe this in terms of a simple damping coefficient, $\gamma$, and an associated stochastic or Langevin force \cite{Langevin}, $F(t)$, which models the effect of environmental fluctuations on the oscillator \cite{Uhlenbeck1930,Wang1945,Risken,Lemons,Mazo}.  The dynamics is described by a simple linear differential equation of the form
\begin{equation}
\label{Eq1}
\ddot{x} + \gamma\dot{x} + \omega^2_0x = \frac{F(t)}{m} ,
\end{equation}
where $m$ is the mass of the oscillating particle.  There is no requirement for detailed knowledge of the fluctuating force, which may be considered to have a very short correlation time with a magnitude determined by the requirements of thermodynamic equilibrium.

The damped quantum harmonic oscillator requires that explicit account be taken of the quantum nature of the environmental degrees of freedom \cite{Senitzky}, which are most simply described by an ensemble of harmonic oscillators \cite{Louisell1964}.  If the damping is very weak, so that $\gamma \ll \omega_0$, then we can neglect rapidly oscillating terms in the coupling between the oscillator and the environmental oscillators by making the rotating wave approximation, which corresponds to enforcing the conservation of the total number of vibrational quanta, and then the Born and Markov approximations associated with weak coupling and loss of memory in the reservoir \cite{Radmore1997}.  This leads to the master equations and Heisenberg-Langevin equations that are ubiquitous, most especially, in quantum optics \cite{Radmore1997,Louisell1973,Grabert1982,Perina,Meystre1991,Carmichael1993,Walls1994,Carmichael1999,Breuer2002,Gardiner2004,Ficek2004,Wiseman2010,Weiss2012}.  

If the coupling is somewhat stronger then it may not be possible to make the rotating wave approximation and we then need to retain in the Hamiltonian terms that can create or annihilate a pair of quanta, one in the damped oscillator and, at the same time, one in the environment.  This leads to the Caldeira-Leggett model \cite{Philbin2012,Agarwal,UllersmaI,UllersmaII,CalderiaI,CaldeiraII,Haake}, which we describe in the following section, and which has been applied to study a wide variety of quantum open systems \cite{Breuer2002,Weiss2012}. A variant on the model has been applied to the quantum theory of light in dielectric \cite{Huttner1992a,Huttner1992b} and magneto-dielectric media \cite{Kheirandish2006,Kheirandish2008,Kheirandish2009,Amooshahi2009}.  An important complication that seems to be an inevitable consequence working in this strong-coupling regime is the failure of the Markov approximation; attempts to enforce this approximation lead to a master equation that is unphysical in that there exist initial states for which the dynamics leads to negative probabilities \cite{Breuer2002,Munro,Stenholm,Cresser2005,Cresser2006}.  It is possible to derive a master equation but the resulting equation is one that has within it non-trivial time-dependent coefficients \cite{Haake,Hu}.  This time-dependence is a clear signature of the non-Markovian nature of the associated evolution.  It seems that this non-Markovian character is an inevitable feature of the strongly-damped quantum harmonic oscillator.


\section{Hamiltonian for the strongly-damped harmonic oscillator}

\begin{figure}[htbp] 
\centering
\includegraphics[width=10cm]{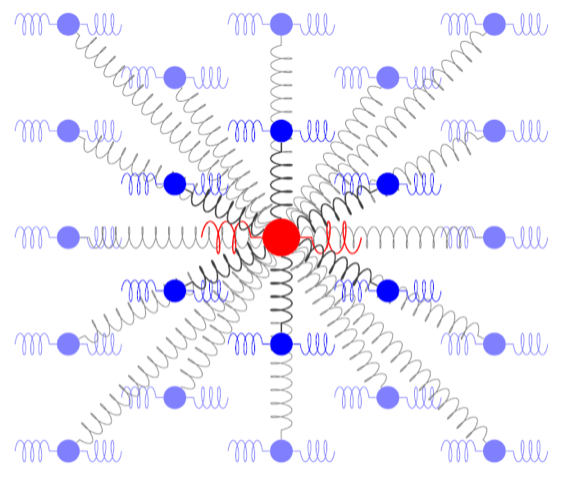}
\caption{Representation of a single harmonic oscillator (in red) coupled harmonically to a bath of 
oscillators (in blue).} 
\label{fig:figure1}
\end{figure}

Consider a harmonic oscillator that is strongly coupled to its environment, ultimately to be treated  as a thermodynamic reservoir, modelled as large collection of oscillators, with a range of frequencies, through their respective positions as depicted in figure \ref{fig:figure1}.  We write the Hamiltonian for the combined oscillator-reservoir system in the form \cite{CalderiaI,Haake}
\begin{equation}
\label{Eq2}
\hat{H} = \frac{\hat{p}^2}{2m} + \frac{1}{2}m\Omega^2_0\hat{x}^2 + 
\sum_\mu\left(\frac{\hat{p}^2_\mu}{2m_\mu} + \frac{1}{2}m_\mu\omega^2_\mu\hat{x}^2_\mu\right)
- \sum_\mu m_\mu\omega_\mu^2\lambda_\mu\hat{x}_\mu\hat{x} .
\end{equation}
If we complete the square we can rewrite this in a minimal-coupling form to arrive at the alternative form:
\begin{equation}
\label{Eq3}
\hat{H} = \frac{\hat{p}^2}{2m} + \frac{1}{2}m\omega^2_0\hat{x}^2 + 
\sum_\mu\left[\frac{\hat{p}^2_\mu}{2m_\mu} + 
\frac{1}{2}m_\mu\omega^2_\mu\left(\hat{x}_\mu - \lambda_\mu\hat{x}\right)^2\right] ,
\end{equation}
where 
\begin{equation}
\label{Eq4}
\omega_0^2 = \Omega_0^2 - \sum_\mu \frac{m_\mu}{m}\omega^2_\mu\lambda^2_\mu .
\end{equation}
Each term in our Hamiltonian (\ref{Eq3}) is strictly positive only if this quantity is positive, corresponding to a real frequency $\omega_0$.  If it is negative then the second term is also negative and the Hamiltonian is not bounded from below and hence not allowed physically.  Hence the positivity of the Hamiltonian places a physical restriction on the strength of the coupling
\begin{equation}
\label{Eq5}
\Omega_0^2 > \sum_\mu \frac{m_\mu}{m}\omega^2_\mu\lambda^2_\mu ,
\end{equation}
while $\omega_0$ can take any positive value.  \footnote{In the literature the change from the Hamiltonian (\ref{Eq2}) to (\ref{Eq3}) is often described as adding a counter term \cite{Breuer2002,Grabert1984}.  We find no need to speak of adding counter terms as both of the natural frequencies $\omega_0$ and $\Omega_0$ play a role in the dynamics.}

At this point it is necessary to pause and consider the fact that our damped harmonic oscillator seems to have \emph{two} possible natural frequencies, $\Omega_0$ and $\omega_0$.  These appear as the potential energy in our two forms of the Hamiltonian given in Eqs. (\ref{Eq2}) and (\ref{Eq3}).  It is reasonable, therefore, to ask which of these (if either) corresponds to the `true' natural frequency. In order to address this important point, we step aside from our principal objective of diagonalizing the Hamiltonian to derive the Heisenberg-Langevin equation for the oscillator position operator.

\subsection{Heisenberg-Langevin equation: a tale of two frequencies}

We find that \emph{both} the frequencies $\omega_0$ and $\Omega_0$ have roles to play in the dynamics of the oscillator and that, in this sense, both fulfil the roles of natural frequency of the damped oscillator, albeit in different time domains.  To demonstrate this we can derive form our Hamiltonian a Heisenberg-Langevin equation of motion for the position operator of our oscillator.  The details are given in \ref{AppHeisenberg}. We find 
\begin{equation}
\label{EqX1}
\ddot{\hat{x}}(t) + \int_0^t \kappa(t-t') \dot{\hat{x}}(t') dt' + \left(\Omega_0^2 - \kappa(0)\right) \hat{x}(t) 
+ \kappa(t)\hat{x}(0) = \frac{\hat{F}(t)}{m} \, ,
\end{equation}
where $\kappa(t)$ is the memory kernel:
\begin{equation}
\label{EqX2}
\kappa(t) = \sum_\mu \frac{m_\mu}{m}\omega_\mu^2\lambda_\mu^2 \cos(\omega_\mu t)
\end{equation}
and $F(t)$ is the Langevin force:
\begin{equation}
\label{EqX3}
\hat{F}(t) = \omega^2_\mu\lambda_\mu\left(\hat{x}_\mu(0) \cos(\omega_\mu t) + \frac{\hat{p}_\mu(0)}{m_\mu\omega_\mu}\sin(\omega_\mu t)\right) \, .
\end{equation} 
Note that this equation of motion is (essentially) exact.  In order to interpret the various terms it suffices to consider the expectation value of this operator equation:
\begin{equation}
\label{EqX4}
\langle\ddot{\hat{x}}(t)\rangle + \int_0^t \kappa(t-t') \langle\dot{\hat{x}}(t')\rangle dt' + \left(\Omega_0^2 - \kappa(0)\right) \langle\hat{x}(t)\rangle 
+ \kappa(t)\langle\hat{x}(0)\rangle = 0 \, ,
\end{equation}
and note that we can also write this in the form
\begin{equation}
\label{EqX5}
\langle\ddot{\hat{x}}(t)\rangle + \int_0^t \kappa(t-t') \langle\dot{\hat{x}}(t')\rangle dt' + \omega_0^2 \langle\hat{x}(t)\rangle 
+ \kappa(t)\langle\hat{x}(0)\rangle = 0 \, .
\end{equation}
The first of these equations is written in terms of the frequency $\Omega_0$ and the second in terms of $\omega_0$. It remains to determine the physical role of each of these, which we can do by considering very short and longer time scales.

\subsubsection{Ultra-short time scales}

To better appreciate what happens at very short times, we first undo the integration by parts that led to our equation of motion and write this in the form
\begin{equation}
\label{EqX6}
\langle\ddot{\hat{x}}(t)\rangle + \int_0^t \dot{\kappa}(t-t') \langle\hat{x}(t')\rangle dt' + \Omega_0^2 \langle\hat{x}(t)\rangle = 0 \, .
\end{equation}
For a very short time, $\delta t$, this becomes
\begin{equation}
\label{EqX7}
\langle\ddot{\hat{x}}(\delta t)\rangle + \Omega_0^2\langle\hat{x}(\delta t)\rangle + O(\delta t^2) = 0 \, .
\end{equation}
The integral term in Eq. (\ref{EqX6}) is of order $\delta t^2$ by virtue of Eq. (\ref{EqX2}) as for short times $\dot{\kappa}$ is of order $\delta t$.  The combination of Eq. (\ref{EqX7}) and $\dot{\hat{x}} = \hat{p}/m$, which is always true, leads to the ultra-short time behaviour
\begin{eqnarray}
\label{EqX8}
\langle\hat{x}(\delta t)\rangle &=& \langle\hat{x}(0)\rangle + \frac{\langle\hat{p}(0)}{m}\delta t \nonumber \\
\langle\hat{p}(\delta t)\rangle &=& \langle\hat{p}(0)\rangle - m\Omega_0^2\langle\hat{x}(0)\rangle\delta t \, .
\end{eqnarray} 
This is the short-time behaviour of a harmonic oscillator of frequency $\Omega_0$.  It is clear that this is the frequency of the oscillator in the non-Markovian regime.

\subsubsection{Longer time scales}

Our first task is to firm up what we mean by longer time.  To do so we note that the function $\kappa (t)$ involves a summation of oscillating cosines of different frequencies, one for each environment oscillator, and that these will dephase, causing $\kappa(t)$ to decay.  We define the longer-time regime as that for which we can approximate $\kappa(t) \approx 0$.  In this regime, our 
our equation of motion (\ref{EqX5}) becomes
\begin{equation}
\label{EqX9}
\langle \ddot{x}(t)\rangle + \int_0^t \kappa(t-t') \langle\dot{x}(t')\rangle dt' + \omega^2_0 \langle x(t)\rangle = 0  \, .
\end{equation}
This equation retains the possibility of non-Markovian effects in the (second) damping term but it is clear that the natural frequency of the oscillator in this time regime is $\omega_0$ and \emph{not} $\Omega_0$.  Thus the question of whether our Hamiltonian applies in the over-damped regime depends on the relationship between the damping and $\omega_0$ and not $\Omega_0$.

We are now in a position to address the issue of whether our model Hamiltonian can be applied in the over-damped regime. To enter this regime, we need to reduce the natural frequency of the oscillator so that it is below the damping rate.  It is clear from the inequality (\ref{Eq5}) that there is always a lower bound below which $\Omega_0$ cannot be reduced.  \emph{However} there is no such bound on $\omega_0$ and, indeed, we can set $\omega_0 = 0^+$ without invalidating our model.  As it is $\omega_0$ and not $\Omega_0$ that is the natural frequency of the oscillator beyond the ultra-short time regime, it is clear that we can use our model to describe an oscillator in the strongly over-damped regime.

A simple example might help to illustrate the ideas presented above.  We consider the simplest case of Ohmic damping, or Ohmic friction \cite{Weiss2012}.  To this end, we go to the continuum limit in which 
\begin{equation}
\label{EqX10}
\kappa(t) \rightarrow \frac{2}{\pi m}\int_0^\infty \frac{J(\omega)}{\omega}\cos(\omega t) d\omega \, ,
\end{equation}
with
\begin{equation}
\label{EqX11}
J(\omega) = m\gamma \omega e^{-\omega/\omega_c} \, .
\end{equation}
Note that it is essential to include a frequency cut-off in this case as, without this, we cannot satisfy the inequality (\ref{Eq5}). We find that 
\begin{equation}
\label{EqX12}
\kappa(t) = \frac{2}{\pi} \frac{\gamma\omega_c}{1 + \omega_c^2t^2} \quad \Rightarrow \quad \kappa(0) = \frac{2\gamma\omega_c}{\pi} \, .
\end{equation}
This also demonstrates the necessity of a cut-off frequency.

Recall that the ultra-short time regime corresponds to times for which we retain the term $\kappa(t)\langle x(0)\rangle$ in (\ref{EqX4}).  This means times for which $t < \omega_c^{-1}$, which fits with the familiar idea that the time needs to be short compared inverse bandwidth of the reservoir.  As there needs to be, in the exact theory, a short-time non-Markovian regime, we need the cut-off frequency in our bath coupling.

For longer times, for which $\kappa(t) \approx 0$, our equation for the expectation value of  the position reduces to (\ref{EqX9}). If the coupling to the reservoir is sufficiently weak that the expectation value $\langle\dot{x}\rangle$ does not vary significantly on the timescale $\omega_0^{-1}$, then we can make the approximation 
\begin{eqnarray}
\label{EqX13}
 \int_0^t \kappa(t-t') \langle\dot{x}(t')\rangle dt' &\approx & \langle\dot{x}(t)\rangle \int_0^t \kappa(t-t')  dt' \nonumber \\
 &=& \gamma \langle\dot{x}(t)\rangle .
\end{eqnarray}
It then follows that our equation for the expectation value of the position becomes (in this Markovian, longer-time regime)
\begin{equation}
\label{EqX14}
\langle \ddot{x}(t)\rangle + \gamma\langle\dot{x}(t)\rangle + \omega^2_0 \langle x(t)\rangle = 0  \, ,
\end{equation}
which is the familiar equation for a damped harmonic oscillator.  Note however, that it is the frequency $\omega_0$ and \emph{not} $\Omega_0$ that appears.  We enter the over-damped regime when $\omega_0 < \gamma/2$.  The constraints on the natural frequencies in (\ref{Eq4}) correspond in this Ohmic damping example to
\begin{equation}
\label{EqX15}
\omega_0^2 > 0 \quad {\rm and} \quad \Omega_0^2 > \kappa(0) = \frac{2\gamma\omega_c}{\pi} \, .
\end{equation}  
We see that there is no lower bound on $\omega_0$ (although it must be real and greater than zero) but there is a bound on $\Omega_0$: it must exceed the geometric mean of the damping rate $\gamma$ and the cut-off frequency $\omega_c$ multiplied by $2/\pi$.  Thus we see again that we cannot allow the cut-off frequency to tend to infinity.  Similarly the frequency $\Omega_0$ will lie somewhere between $\gamma$ and the rather larger $\omega_c$ so we cannot have $\Omega_0$ less than $\gamma$.  What saves the model in the strongly damped regime is the fact that it is $\omega_0$ rather than $\Omega_0$ that corresponds to the natural frequency of the damped oscillator.

As a final note in this section, we note that the presence of at least two candidate natural frequencies is all but inevitable for a strongly damped oscillator, as such system will, in general, always experience a non-Markovian short-time evolution.  As we shall see, the existence of these two frequencies also complicates the question of the amount of energy associated with the oscillator during its evolution, but especially in its steady state.


\subsection{Continuum reservoir}

Our first, perhaps, less familiar feature is to replace the discrete reservoir of oscillators by a continuum. To proceed, we first rewrite our Hamiltonian in terms of the familiar annihilation and creation operators:
\begin{eqnarray}
\label{Eq6}
\hat{a} = \sqrt{\frac{m\Omega_0}{2\hbar}}\left(\hat{x} + \frac{i\hat{p}}{m\Omega_0}\right)  \nonumber \\
\hat{b}_\mu = \sqrt{\frac{m_\mu\omega_\mu}{2\hbar}}\left(\hat{x}_\mu + \frac{i\hat{p}_\mu}{m_\mu\omega_\mu}\right)
\end{eqnarray}
In terms of these operators our Hamiltonian, Eq. (\ref{Eq2}), becomes
\begin{equation}
\label{Eq7}
\hat{H} = \hbar\Omega_0\hat{a}^\dagger\hat{a} + \sum_\mu\hbar\omega_\mu\hat{b}_\mu^\dagger\hat{b}_\mu
+ \sum_\mu\frac{\hbar}{2} V_\mu\left(\hat{a} + \hat{a}^\dagger\right)\left(\hat{b}_\mu + \hat{b}^\dagger_\mu\right)
\end{equation}
when unimportant constant shifts in the ground-state energies are removed and
\begin{equation}
\label{Eq8}
V_\mu = - \sqrt{\frac{m_\mu\omega_\mu}{m\Omega_0}}\omega_\mu\lambda_\mu .
\end{equation}
When written in terms of this quantity, our positivity condition (\ref{Eq5}) becomes
\begin{equation}
\label{Eq9}
\Omega_0 > \sum_\mu \frac{V^2_\mu}{\omega_\mu} .
\end{equation}
At this stage we can seek to diagonalise the full Hamiltonian to find, in effect, the normal modes of the oscillator coupled to the reservoir.  This is the approach taken by Haake and Reibold and by Ford \emph{et al} \cite{Haake,O'Connell}.  The dynamics are then reminiscent of the Bixon-Jortner model, with recurrences occurring on a timescale given by the inverse of the frequency spacing of the reservoir oscillators \cite{Radmore1997,Bixon,Kyrola,Milonni,Eberly,Tarzi}, as is characteristic of periodic and almost periodic functions \cite{Bohr} \footnote{As is often the case in science, the Bixon-Jortner model itself had an anticipation in the early work of Fano \cite{Fano35}. We are grateful to Jan Petter Hansen for bringing Fano's paper to our attention.}.

We find it both simpler and also more powerful to first recast our model in terms of a continuum description of the reservoir. 
To this end we introduce continuum annihilation and creation operators, $\hat{b}(\omega)$ and $\hat{b}^\dagger(\omega)$, satisfying the commutation relations
\begin{equation}
\label{Eq10}
\left[\hat{b}(\omega), \hat{b}^\dagger(\omega')\right] = \delta(\omega - \omega') 
\end{equation}
and our Hamiltonian becomes \cite{Huttner1992b}
\begin{eqnarray}
\label{Eq11}
\hat{H} = \hbar\Omega_0\hat{a}^\dagger\hat{a} + \int_0^\infty d\omega \:\hbar\omega \hat{b}^\dagger(\omega)\hat{b}(\omega) \nonumber \\
  \qquad \qquad +\int_0^\infty d\omega \: \frac{\hbar}{2}V(\omega)\left(\hat{a} + \hat{a}^\dagger\right)\left[\hat{b}^\dagger(\omega)
+\hat{b}(\omega)\right] 
\end{eqnarray}
and the positivity condition is then
\begin{equation}
\label{Eq12}
\Omega_0 > \int_0^\infty d\omega \frac{V^2(\omega)}{\omega} \, .
\end{equation}
Our Hamiltonian is quadratic in the annihilation and creation operators for the oscillator and the reservoir and hence leads to linear coupled equations of motion for these operators.  We could seek to solve these equations of motion and this would lead to an operator Heisenberg-Langevin equation similar to that derived above and in a number of earlier texts \cite{Grabert1984,Ford1985,Ford1985a}.  Here we adopt the different approach of diagonalising the Hamiltonian.

\subsection{Hamiltonian diagonalisation}

Our second less familiar element is the exact diagonalisation of the oscillator-reservoir Hamiltonian. We shall find that this can be achieved with greater generality than is possible for the model with discrete reservoir oscillators simply because there is greater freedom in the evaluation of integrals than summations. Our task is to diagonalise the Hamiltonian by finding a complete set of eigenoperators, $\hat{B}(\omega)$ and their conjugates $\hat{B}^\dagger(\omega)$ that satisfy the operator equations
\begin{eqnarray}
\label{Eq13}
\left[\hat{B}(\omega), \hat{H}\right] = \hbar\omega \hat{B}(\omega) \nonumber \\
\left[\hat{B}^\dagger(\omega), \hat{H}\right] = -\hbar\omega \hat{B}^\dagger(\omega)
\end{eqnarray}
for all positive frequencies $\omega$.  These eigenoperators are complete if, in addition to these they also satisfy the condition
\begin{equation}
\label{Eq14}
\left[\hat{B}(\omega),\hat{B}^\dagger(\omega')\right] = \delta(\omega-\omega')  .
\end{equation}
These operator equations are the natural analogues of the more familiar eigenvalue and completeness conditions for the eigenstates of a Hamiltonian \cite{Radmore1997,Radmore1988}.  In analogy with the eigenvalue problem, we expand each of the eigenoperators as a superposition of a complete set of operators:
\begin{equation}
\label{Eq15}
\hat{B}(\omega) = \alpha(\omega)\hat{a} + \beta(\omega)\hat{a}^\dagger + 
\int_0^\infty d\omega' \: \left[\gamma(\omega,\omega')\hat{b}(\omega') + \delta(\omega,\omega')\hat{b}^\dagger(\omega')\right] ,
\end{equation}
and then use the eigenoperator equations and completeness condition to determine the coefficients in this expansion.  The calculation is a little involved but the main points are summarised in \ref{Fano}.

We can express any of the annihilation and creation operators for the oscillator or the reservoir in terms of our eigenoperators.  To do this we write the desired operator as a superposition of all the $\hat{B}(\omega)$ and $\hat{B}^\dagger(\omega)$ operators and then use the commutation relations to extract the coefficients in this expansion.  For the oscillator operators we find
\begin{eqnarray}
\label{Eq16}
\hat{a} = \int_0^\infty d\omega \left(\alpha^*(\omega)\hat{B}(\omega) - \beta(\omega)\hat{B}^\dagger(\omega)\right)
\nonumber \\
\hat{a}^\dagger = \int_0^\infty d\omega \left(\alpha(\omega)\hat{B}^\dagger(\omega) - \beta^*(\omega)\hat{B}(\omega)\right) .
\end{eqnarray}
The requirement that these operators satisfy the familiar boson commutation relation, $\left[\hat{a},\hat{a}^\dagger\right] = 1$, provides a constraint on the functions $\alpha(\omega)$ and $\beta(\omega)$ in the form
\begin{equation}
\label{Eq17}
\int_0^\infty d\omega \left[|\alpha(\omega)|^2 - |\beta(\omega)|^2\right] = 
\int_0^\infty d\omega |\alpha(\omega)|^2\frac{4\Omega_0\omega}{(\Omega_0 + \omega)^2} = 1 ,
\end{equation}
where we have used (\ref{EqA10}).  It is interesting to note that the correctness of this may be verified explicitly by contour integration and that the proof makes explicit use of the positivity condition (\ref{Eq12}) \cite{Huttner1992b}.

The integrand in (\ref{Eq17}) is clearly positive (or zero) for all frequencies and is also normalized, and hence it has the mathematical form of a frequency probability distribution:
\begin{equation}
\label{Eq18}
\pi(\omega) =  |\alpha(\omega)|^2\frac{4\Omega_0\omega}{(\Omega_0 + \omega)^2}  .
\end{equation} 
A number of further constraints on this quantity emerge naturally from thinking of it as a probability distribution and from our diagonalization. We note that the same concept has been described before: Georgievskii and Pollak introduce an effective density of states for the diagonalized Caldeira-Leggett model \cite{Georgievskii}, while Ratchov \emph{et al} have expressed the properties of the damped harmonic oscillator in terms of such a probability density \cite{Ratchov}, see also \cite{Ghosh}.

\subsection{Physical constraints}

We leave until the next section a discussion of the physical interpretation of the probability density $\pi(\omega)$ but consider here what can be inferred from the fact that it has the mathematical properties of a probability distribution.  To this end let us denote the average value of a function of frequency for this distribution by
\begin{equation}
\label{Eq19}
\langle\langle f(\omega) \rangle\rangle = \int_0^\infty d\omega f(\omega)\pi(\omega) .
\end{equation}
we note that $|\alpha(\omega)|^2$ is finite for all $\omega$ and it follows, therefore, from (\ref{Eq18}) that the average value $\langle\langle\omega^{-1}\rangle\rangle$ is finite.

It follows from the eigenoperator equations (\ref{Eq13}) that we can write the Hamiltonian in the form
\begin{equation}
\label{Eq20}
\hat{H} = \int_0^\infty d\omega \:\hbar\omega\: \hat{B}^\dagger(\omega)\hat{B}(\omega) + C  ,
\end{equation}
where $C$ is an unimportant constant (even though it is formally divergent).  We can substitute into this Hamiltonian our expression for the eigenoperators (\ref{Eq15}).  We require that the coefficients of $\hat{a}^2$ and $\hat{a}^{\dagger2}$ should vanish so that
\begin{equation}
\label{Eq21}
\int_0^\infty d\omega \:\omega |\alpha(\omega)|^2\left(\frac{\omega - \Omega_0}{\omega + \Omega_0}\right) = 
\int_0^\infty d\omega \frac{\pi(\omega)}{4\Omega_0}(\omega^2 - \Omega_0^2)
= 0 \, ,
\end{equation}
where we have again used (\ref{EqA10}).  This implies that 
\begin{equation}
\label{Eq22}
\langle\langle \omega^2 \rangle \rangle = \Omega^2_0 .
\end{equation}
The fact that the square of the mean value cannot exceed the mean value of the square for any probability distribution leads us to deduce that
\begin{equation}
\label{Eq23}
\langle\langle\omega\rangle\rangle < \Omega_0 .
\end{equation}

Finally we can apply the Cauchy-Schwartz inequality to provide a lower bound on the value of 
$\langle\langle\omega^{-1}\rangle\rangle$:
\begin{eqnarray}
\label{Eq24}
\langle\langle\omega^{-1}\rangle\rangle \langle\langle\omega\rangle\rangle \ge  1 
\nonumber \\
\Rightarrow \quad \langle\langle\omega^{-1}\rangle\rangle > \frac{1}{\Omega_0} .
\end{eqnarray}
These inequalities are useful in determining the properties of the ground state.


\section{Ground-state}
It is immediately clear from the form of the Hamiltonian (\ref{Eq11}) that the ground state of the oscillator is not that of the undamped oscillator, that is the state annihilated by $\hat{a}$.  To see this we need only note that the interaction term $\hat{a}^\dagger\hat{b}^\dagger(\omega)$ acts on the combined ground state of the non-interacting oscillator and reservoir, adding a quantum to each.  This suggests that the true ground state should be a superposition of states with varying numbers of quanta in both the oscillator and the reservoir and hence an entangled state.  This situation is reminiscent of the ground state of an atom in quantum electrodynamics, which is dressed by virtual photons \cite{Franco}.  The dressing of the ground-state atom is responsible for a number of important effects including the Casimir-Polder interaction \cite{Casimir,Edwin,Craig} and the form of the polarizability of the atom \cite{Rodney2006,Milonni2008}.It is reasonable to expect that it will be similarly significant for our strongly damped oscillator.  

The true ground state, which we denote by the ket $|0\rangle$, is the zero-eigenvalue eigenstate of all the annihilation operators $\hat{B}(\omega)$:
\begin{equation}
\label{Eq25}
\hat{B}(\omega)|0\rangle = 0 \qquad \qquad \forall \: \omega .
\end{equation} 
The ground-state properties of the oscillator in this pure state are described by a mixed state density operator obtained by tracing out the environmental degrees of freedom:
\begin{equation}
\label{Eq26}
\hat{\rho}_{\rm Osc} = {\rm Tr}_{\rm Env}\left(|0\rangle\langle 0|\right) .
\end{equation}
The most straightforward way to determine the form of this mixed state is to use the characteristic function \cite{Radmore1997}:
\begin{eqnarray}
\label{Eq27}
\chi(\xi) &=& {\rm Tr}\left[\hat{\rho}\exp\left(\xi\hat{a}^\dagger - \xi^*\hat{a}\right)\right] \nonumber \\
&=& \langle 0|\exp\left(\xi\hat{a}^\dagger - \xi^*\hat{a}\right)|0\rangle .
\end{eqnarray}
This function provides a complete description of the state and all of its statistical properties.  A brief summary of the principal properties of this characteristic function is given in \ref{Characteristic}. To evaluate the characteristic function it we express $\hat{a}$ and $\hat{a}^\dagger$ in terms of the eigenoperators\footnote{We make use of the operator ordering theorem \cite{Radmore1997}
\begin{eqnarray}
\nonumber
\exp(\hat{A}+\hat{B}) = \exp(\hat{A})\exp(\hat{B})
\exp\left(-\frac{1}{2}[\hat{A},\hat{B}]\right) ,
\end{eqnarray}
which holds if the two operators $\hat{A}$ and $\hat{B}$ both commute with their commutator 
$[\hat{A},\hat{B}]$.}
\begin{eqnarray}
\label{Eq28}
\chi(\xi) &=& \langle 0|\exp\left\{\int_0^\infty  d\omega
\left[\left(\xi\alpha+\xi^*\beta\right)\hat{B}^\dagger(\omega)
- \left(\xi^*\alpha^*+\xi\beta^*\right)\hat{B}(\omega)
\right]\right\}|0\rangle \nonumber \\
&=& \exp\left(-\frac{1}{2}\int_0^\infty d\omega|\xi\alpha(\omega)+\xi^*\beta(\omega)|^2\right) .
\end{eqnarray}
This simple form, Gaussian in $\xi$, is characteristic of a squeezed thermal state.  We can rewrite this characteristic function in terms of our probability density in the form
\begin{eqnarray}
\label{Eq29}
\chi(\xi) &=& \exp\left[-\frac{1}{2}\int_0^\infty d\omega \: \pi(\omega)\left(\frac{\omega}{\Omega_0}\xi_r^2
+ \frac{\Omega_0}{\omega}\xi_i^2\right)\right] \nonumber \\
&=& \exp\left[-\frac{1}{2}\left(\frac{\langle\langle\omega\rangle\rangle}{\Omega_0}\xi_r^2
+ \langle\langle\omega^{-1}\rangle\rangle\Omega_0\:\xi_i^2\right)\right] ,
\end{eqnarray}
where $\xi_r$ and $\xi_i$ are the real and imaginary parts of $\xi$ respectively.  We note, in passing, that the quadrature operators for the oscillator (familiar from quantum optics \cite{Radmore1997}) have unequal uncertainties:
\begin{eqnarray}
\label{Eq30}
\Delta\left(\frac{\hat{a}+\hat{a}^\dagger}{\sqrt{2}}\right) = \sqrt{\frac{\langle\langle\omega^{-1}\rangle\rangle\Omega_0}{2}}
\nonumber \\
\Delta\left(\frac{-i(\hat{a}-\hat{a}^\dagger)}{\sqrt{2}}\right) = 
\sqrt{\frac{\langle\langle\omega\rangle\rangle}{2\:\Omega_0}}  .
\end{eqnarray}
The product of these two exceeds $\frac{1}{2}$, as it must, by virtue of the Cauchy-Schwartz inequality (\ref{Eq24}).

\subsection{Ground-state energy}

The ground-state energy of the undamped harmonic oscillator is simply $\frac{1}{2}\hbar\omega_0$, where $\omega_0$ is the natural angular frequency of the oscillator.  For an underdamped oscillator, the oscillation frequency is $\sqrt{\omega_0^2 - \gamma^2/4}$, which is less than the frequency of the undamped oscillator and this suggests, perhaps, that the ground-state energy is similarly reduced \cite{Philbin2013}. Although plausible, this would clearly run into difficulties in the over-damped regime in which this characteristic frequency becomes imaginary.  We shall see that the ground-state energy of the oscillator, however we define it, \emph{increases} as a consequence of the damping.

We have established that in the ground state, the reduced state of the oscillator is Gaussian in position and momentum and, as such, is completely determined by the first and second moments of the position and momentum.  These moments are
\begin{eqnarray}
\label{Eq31}
\langle 0|\hat{x}|0\rangle = 0 \nonumber \\
\langle 0|\hat{p}|0\rangle = 0 \nonumber \\
\langle 0|\hat{x}^2|0\rangle = \frac{\hbar\langle\langle\omega^{-1}\rangle\rangle}{2m} \nonumber \\
\langle 0|\hat{p}^2|0\rangle = \frac{\hbar m\langle\langle\omega\rangle\rangle}{2} \nonumber \\
\langle 0|\hat{p}\hat{x} + \hat{p}\hat{x}|0\rangle = 0 .
\end{eqnarray}
Expressions of this form were derived previously by Grabert and Weiss in their theory of the damped harmonic oscillator \cite{Grabert1984}. We note that the frequency of the undamped oscillator, be it $\omega_0$ or $\Omega_0$, makes no explicit appearance in these expectation values, but the averages $\langle\langle\omega\rangle\rangle$ and $\langle\langle\omega^{-1}\rangle\rangle$ will depend on these frequencies.

We have seen that the quantum damped harmonic oscillator is characterised by two natural frequencies, $\omega_0$ and $\Omega_0$, and it is natural to define the energy of the oscillator in terms of one or other of these:
\begin{equation}
\label{Eq31a}
\hat{H}_{f_0} = \frac{\hat{p}^2}{2m} + \frac{1}{2}m f^2_0\hat{x}^2 \, ,
\end{equation}
where $f_0 = \omega_0$ or $\Omega_0$.  We shall consider a third possible natural frequency, $\omega_{\rm diag}$, and the associated energy below.  The ground-state expectation values, Eq. (\ref{Eq31}), mean that we can write the expectation value of the energy in the form
\begin{equation}
\label{Eq31b}
\langle 0|\hat{H}_{f_0}|0\rangle = \frac{\hbar f_0}{4}\left(\frac{\langle\langle\omega\rangle\rangle}{f_0} + 
\langle\langle\omega^{-1}\rangle\rangle f_0\right) \, .
\end{equation}
Recall that the Cauchy-Schwartz inequality requires that
\begin{equation}
\label{Eq31c}
\langle\langle\omega^{-1}\rangle\rangle \langle\langle\omega\rangle\rangle \geq 1
\quad \Rightarrow \langle\langle\omega^{-1}\rangle\rangle \geq \frac{1}{\langle\langle\omega\rangle\rangle}
\end{equation}
and it follows that the ground-state energy is bounded by 
\begin{equation}
\label{Eq31d}
\langle 0|\hat{H}_{f_0}|0\rangle \geq \frac{\hbar f_0}{4}\left(\frac{\langle\langle\omega\rangle\rangle}{f_0} + 
\frac{f_0}{\langle\langle\omega\rangle\rangle}\right) \, .
\end{equation}
The global minimum of this expression occurs for $\langle\langle\omega\rangle\rangle = f_0$ and hence 
\begin{equation}
\label{Eq31e}
\langle 0|\hat{H}_{f_0}|0\rangle \geq \frac{1}{2}\hbar f_0
\end{equation}
for \emph{any} choice of frequency in the expression for the oscillator energy:
\begin{eqnarray}
\label{Eq31f}
\langle 0|\hat{H}_{\Omega_0}|0\rangle \geq \frac{1}{2}\hbar\Omega_0  \nonumber \\
\langle 0|\hat{H}_{\omega_0}|0\rangle \geq \frac{1}{2}\hbar\omega_0  \, .
\end{eqnarray}
The fact that both of these exceed
the ground-state energy of the corresponding undamped oscillator is a reflection of the fact that there is an energy cost to be paid in order to decouple the oscillator from its environment \cite{Allahverdyan}.  This increase conflicts, however, with the reduction in ground-state energy that has previously been reported \cite{Philbin2013}. We note, further, that the mean kinetic energy and potential energy for the oscillator alone do not have the same value and that this is in marked contrast to the ground state of the undamped oscillator.  

\subsection{Diagonal form of the oscillator ground state}

We can diagonalise the density operator for the oscillator alone, $\hat{\rho}_{\rm Osc}$, by means of a squeezing transformation \cite{Radmore1997} or, equivalently, introducing a new pair of annihilation and creation operators corresponding to a third candidate natural oscillation frequency, $\omega_{\rm diag}$:
\begin{eqnarray}
\label{Eq33}
\hat{c} = \sqrt{\frac{m\omega_{\rm diag}}{2\hbar}}\left(\hat{x} + i\frac{\hat{p}}{m\omega_{\rm diag}}\right) 
\nonumber \\
\hat{c}^\dagger = \sqrt{\frac{m\omega_{\rm diag}}{2\hbar}}\left(\hat{x} - i\frac{\hat{p}}{m\omega_{\rm diag}}\right) .
\end{eqnarray}
To complete the diagonalisation we need only choose $\omega_{\rm diag}$ such that the expectation values of $\hat{c}^2$ and $\hat{c}^{\dagger 2}$ are zero:
\begin{eqnarray}
\label{Eq34}
\langle 0|\hat{c}^2|0\rangle = \frac{m\omega_{\rm diag}}{2\hbar}
\langle 0|\hat{x}^2 - \frac{\hat{p}^2}{m^2\omega_{\rm diag}^2}|0\rangle  = 0 \nonumber \\
\Rightarrow \omega_{\rm diag} = \sqrt{\frac{\langle\langle\omega\rangle\rangle}{\langle\langle\omega^{-1}\rangle\rangle}} .
\end{eqnarray}
This frequency is the geometric mean of the two frequencies $\langle\langle\omega\rangle\rangle$ and $\langle\langle\omega^{-1}\rangle\rangle^{-1}$ and it is, by virtue of (\ref{Eq23}) and (\ref{Eq24}), less than $\Omega_0$.  


The mean number of $c$-quanta in the oscillator ground-state is
\begin{eqnarray}
\label{Eq35}
\bar{n}_c = \langle 0|\hat{c}^\dagger\hat{c}|0\rangle = 
\frac{1}{2}\left(\sqrt{\langle\langle\omega\rangle\rangle\langle\langle\omega^{-1}\rangle\rangle} - 1\right) ,
\end{eqnarray}
which exceeds $0$, as it should, by virtue of (\ref{Eq24}).  When written in terms of the $c$-quanta, the steady-state density operator takes the form of a thermal Bose-Einstein state, which we can write in the form
\begin{equation}
\label{Eq36}
\hat{\rho}_{\rm Osc} = \frac{1}{\bar{n}_c + 1}\left(\frac{\bar{n}_c}{\bar{n}_c + 1}\right)^{\hat{c}^\dagger\hat{c}} .
\end{equation}
We may interpret this state as a thermal state for the oscillator at the shifted frequency $\omega_{\rm diag}$ and at an effective ``temperature''
\begin{equation}
\label{Eq37}
T_{\rm eff} = \frac{\hbar\omega_{\rm diag}}{k_B\ln(1 + \bar{n}_c^{-1})} .
\end{equation}
We should note, however, that the true temperature in the ground-state is zero and that this quantity and the frequency $\omega_{\rm diag}$ are at most only parameters with which to quantify the state of the oscillator and its entanglement with the environment.  In particular, the von-Neumann entropy associated with the steady state of the oscillator is
\begin{equation}
\label{Eq38}
S({\rm Osc}) = (\bar{n}_c + 1)\ln(\bar{n}_c + 1) - \bar{n}_c \ln\bar{n}_c \, .
\end{equation}
By virtue of the Araki-Lieb inequality \cite{Araki,Wehrl,QIbook} and the fact that the full state is pure, this means that this is also the total entropy of the environment:
\begin{equation}
\label{Eq39}
S({\rm Env}) = S({\rm Osc}) 
\end{equation}
and that the quantum mutual information \cite{QIbook}, or index of correlation \cite{Simon}, between the oscillator and the environment is
\begin{equation}
\label{Eq40}
S({\rm Osc}:{\rm Env}) = S({\rm Env}) + S({\rm Osc}) - S({\rm Osc},{\rm Env}) = 2S({\rm Osc}) .
\end{equation}


\section{Physical meaning of $\pi(\omega)$}

We have seen that the physical properties of the oscillator ground state may readily be expressed in terms of moments of the frequency given the probability distribution $\pi(\omega)$.  Here we present the case for identifying this probability density with the contribution from the dressed modes, associated with the eigenoperators $\hat{B}(\omega)$, to the state of the oscillator.  It is, in essence, in the spectrum of the true ground-state (continuum) modes contributing to the oscillator state.  We present three arguments to support this interpretation.

Our first justification arises from the form of the expectation values (\ref{Eq31}).  We know that the ground state of a harmonic oscillator of frequency $\omega$ has 
\begin{eqnarray}
\label{Eq41}
\langle\hat{x}^2\rangle = \frac{\hbar}{2m\omega} \nonumber \\
\langle\hat{p}^2\rangle = \frac{\hbar m\omega}{2} .
\end{eqnarray}
If we treat the state of the oscillator as a mixture of oscillators of different frequencies, each in its ground state, and contribution with weight $\pi(\omega)$ then the resulting average mean-square values will be
\begin{eqnarray}
\label{Eq42}
\langle\hat{x}^2\rangle = \int_0^\infty d\omega \: \pi(\omega) \frac{\hbar}{2m\omega} 
= \frac{\hbar\langle\langle\omega^{-1}\rangle\rangle}{2m}\nonumber \\
\langle\hat{p}^2\rangle = \int_0^\infty d\omega \: \pi(\omega)\frac{\hbar m\omega}{2} 
= \frac{\hbar m\langle\langle\omega\rangle\rangle}{2} ,
\end{eqnarray}
which correspond to those obtained above.  Note that the requirement that $\langle\hat{x}^2\rangle$ must be finite imposes the condition that at zero frequency
\begin{equation}
\label{Eq42a}
\pi(0) = 0 .
\end{equation}

Our second point arises from the form of the Hamiltonian for the oscillator
\begin{equation}
\label{Eq43}
\hat{H}_{\Omega_0} = \frac{\hat{p}^2}{2m} + \frac{1}{2}m\Omega^2_0\hat{x}^2 .
\end{equation}
We can, by virtue of (\ref{Eq22}), write this as a combination of potentials corresponding to different frequencies but weighted by $\pi(\omega)$:
\begin{eqnarray}
\label{Eq44}
\hat{H}_{\Omega_0} &=& \frac{\hat{p}^2}{2m} + \frac{1}{2}m\int_0^\infty d\omega \: \pi(\omega) \omega^2 \hat{x}^2
\nonumber \\
&=& \frac{\hat{p}^2}{2m} + \frac{1}{2}m \langle\langle\omega^2\rangle\rangle \hat{x}^2 .
\end{eqnarray}

Finally we note that the effective mean energy of the oscillator, which is associated with the diagonal form of the density operator (\ref{Eq36}) is
\begin{equation}
\label{Eq45}
\left(\bar{n}_c + \frac{1}{2}\right)\hbar\omega_{\rm diag} = \frac{1}{2}\hbar\int_0^\infty d\omega \: \pi(\omega) \omega
= \frac{1}{2}\hbar\langle\langle\omega\rangle\rangle ,
\end{equation}
which neatly combines the characteristic ground state energies of the dressed oscillators, weighted by the probability distribution $\pi(\omega)$.  We note that this mean energy is less than $\frac{1}{2}\hbar\Omega_0$ by virtue of Eq (\ref{Eq23}), It is, however, always positive irrespective of whether the oscillator evolution is under- or over-damped.
The general question of whether the ground-state energy of the damped oscillator is greater or less than that of the undamped oscillator is difficult to answer definitively as there is no unique form in the damped oscillator Hamiltonian for the free or undamped oscillator frequency.

The combination of these three features (the expectation values $\langle \hat{x}^2\rangle$ and $\langle \hat{p}^2\rangle$, the form of the oscillator Hamiltonian and the effective mean energy) leads us to interpret $\pi(\omega)$ as the proportion of the corresponding dressed oscillators contributing to the properties of the damped oscillator.  We emphasise that the mathematical results obtained in the preceding section do not \emph{require} us to adopt this interpretation of $\pi(\omega)$ but we find it helpful to do so.

\section{Equilibrium state at finite temperature}

We require the forms of the steady state and the dynamics for our damped harmonic oscillator in an environment at finite temperature.  In evaluating this, we show that the equilibrium state is precisely the mean-force Gibbs state found by tracing out the environmental degrees of freedom from the global equilibrium Gibbs state.  The analysis presents a problem as we cannot assign a thermal state density operator in the continuum limit model of the environment.  We could take a step back and and return to a description in terms of discrete reservoir modes, but it is more natural to adopt the thermofield technique devised for the treatment of problems in finite temperature quantum field theory \cite{Takahashi,Umezawa1982,Knight1985,Dalton1987,Umezawa1993,Dalton2015}.  This is our third less familiar element, and as thermofields will be a novelty for much of intended readership, we present a brief account of the thermofield technique in \ref{thermofields}.

It is \emph{essential} to realise, however, that we have not as yet established that this equilibrium state is also the steady state of the strongly damped oscillator.  We turn to this in the following section.

We can determine the properties of the anticipated steady state by following the same method as employed to study the ground state.  The key idea is to replace the vacuum state \textbf{$|0\rangle$ of the coupled system, see Eq.\ (\ref{Eq25})} with the thermal vacuum state in a doubled space, which is related to the true vacuum state in the doubled space, $|0,\tilde{0}\rangle$, by a unitary transformation:
\begin{eqnarray}
\label{Eq45a}
|0(\beta)\rangle &=& \hat{S}(\theta[\omega])|0,\tilde{0}\rangle 
\nonumber \\
&=& \exp\left[\int d\omega \: \theta(\beta,\omega)\left(\hat{\tilde{B}}^\dagger(\omega) \hat{B}^\dagger(\omega) 
- \hat{B}(\omega)\hat{\tilde{B}}(\omega)\right)\right]|0,\tilde{0}\rangle  \, .
\end{eqnarray}
This state has the same single reservoir expectation values as the thermal state and may therefore be used in its place.  \emph{If} the coupled oscillator-reservoir system relaxes to the thermal state of the coupled system (and we have yet to establish this) then we can use this thermal vacuum state.

The corresponding thermal steady state of the harmonic oscillator, the mean force Gibbs state \cite{Kirkwood,Trushechkin}, will be a mixed state density operator obtained by tracing over the environment:
\begin{equation}
\label{Eq45b}
\hat{\rho}_{{\rm Osc},T} = {\rm Tr}_{\rm Env}[|0(\beta)\rangle\langle 0(\beta)|] \, .
\end{equation}
As with the zero temperature ground state, we can determine the form of this using the characteristic function:
\begin{eqnarray}
\label{Eq45c}
\chi_T(\xi) &=& \langle 0(\beta)|\exp\left(\xi\hat{a}^\dagger - \xi^*\hat{a}\right)|0(\beta)\rangle  \nonumber \\
&=& \langle 0(\beta)|\exp \left\{ \int d\omega \left[ (\xi\alpha+\xi^*\beta)\hat{B}^\dagger(\omega)
- (\xi^*\alpha^* + \xi\beta^*) \hat{B}(\omega) \right] \right\} |0(\beta)\rangle  \, .  \nonumber \\
&& 
\end{eqnarray}
We can transform this into a vacuum expectation value by applying a unitary transformation to the annihilation and creation operators $\hat{B}(\omega)$ and $\hat{B}^\dagger(\omega)$:
\begin{eqnarray}
\label{Eq45d}
\hat{B}(\omega) &\rightarrow& \hat{B}(\omega){\rm cosh}\theta(\beta,\omega) + \hat{\tilde{B}}^\dagger(\omega){\rm sinh}\theta(\beta,\omega)
\nonumber \\
\hat{B}^\dagger(\omega) &\rightarrow& \hat{B}^\dagger(\omega){\rm cosh}\theta(\beta,\omega) + \hat{\tilde{B}}(\omega){\rm sinh}\theta(\beta,\omega) \, .
\end{eqnarray}
Applying this transformation to our characteristic function replaces (\ref{Eq45c}) by an equivalent vacuum expectation value.  Evaluating this gives
\begin{eqnarray}
\label{Eq45e}
\chi_T(\xi) &=& \exp\left(-\frac{1}{2}\int d\omega \: |\xi\alpha(\omega) + \xi^*\beta(\omega)|^2 [{\rm cosh}^2\theta(\beta,\omega)
+ {\rm sinh}^2\theta(\beta,\omega)]\right)  \nonumber \\
&=& \exp\left(-\frac{1}{2}\int d\omega \: |\xi\alpha(\omega) + \xi^*\beta(\omega)|^2 {\rm coth}(\beta\hbar\omega/2)\right) \, .
\end{eqnarray}
As with the zero-temperature steady state, this is a simple Gaussian in $\xi$ and, again, is characteristic of a squeezed thermal state.  When expressed in terms of our probability density, $\pi(\omega)$, we find:
\begin{eqnarray}
\label{Eq45f}
\chi_T(\xi) &=& \exp\left[-\frac{1}{2}\int d\omega \: \pi(\omega){\rm coth}(\beta\hbar\omega/2)\left(
\frac{\omega}{\Omega_0}\xi_r^2 + \frac{\Omega_0}{\omega}\xi_i^2\right)\right]  \nonumber \\
&=& \exp\left[-\frac{1}{2}\left(\frac{\langle\langle \omega {\rm coth}(\beta\hbar\omega/2)\rangle\rangle}{\Omega_0}\xi_r^2
+ \Omega_0\langle\langle\omega^{-1}{\rm coth}(\beta\hbar\omega/2)\rangle\rangle\xi_i^2\right)\right]  \, .  \nonumber \\
& & 
\end{eqnarray}
We note that this has the same general form as the characteristic function for the ground state, Eq (\ref{Eq29}), but with the probability density $\pi(\omega)$ replaced by a thermally-weighted density $\pi(\omega){\rm coth}(\beta\hbar\omega/2)$. With this substitution, we can simply modify the properties of the ground state so that, for example, the lowest moments of the position and momentum operators in this state become
\begin{eqnarray}
\label{Eq45g}
\langle \hat{x}\rangle_T &=& 0   \nonumber \\
\langle \hat{p}\rangle_T &=& 0 \nonumber \\
\langle \hat{x}^2\rangle_T &=& \frac{\hbar\langle\langle \omega^{-1}{\rm coth}(\beta\hbar\omega/2)\rangle\rangle}{2m}  \nonumber \\
\langle \hat{p}^2\rangle_T &=& \frac{\hbar m\langle\langle \omega{\rm coth}(\beta\hbar\omega/2)\rangle\rangle}{2}  \nonumber \\
\langle \hat{x}\hat{p} + \hat{p}\hat{x}\rangle_T &=& 0 .
\end{eqnarray}
We have seen that the oscillator is characterised by (at least) two different natural frequencies, $\Omega_0$ and $\omega_0$.  In terms of these, the mean energy of the oscillator alone is
\begin{eqnarray}
\label{Eq45h}
\left\langle \frac{\hat{p}^2}{2m} + \frac{mf_0^2\hat{x}^2}{2}\right\rangle_T
=  \frac{\hbar f_0}{4}\left(\frac{\langle\langle \omega{\rm coth}(\beta\hbar\omega/2)\rangle\rangle}{f_0}
+ \langle\langle \omega^{-1}{\rm coth}(\beta\hbar\omega/2)\rangle\rangle f_0 \right)  , \nonumber \\
& & 
\end{eqnarray}
where $f_0 = \omega_0$ or $\Omega_0$.  This is the natural generalisation of the ground-state energy of the oscillator given in Eq. (\ref{Eq31b})

It is interesting to pause at this point and to consider the behaviour of the oscillator kinetic and potential energies in the high temperature limit.  For a weakly damped oscillator, we would expect both of these quantities to approach $\frac{1}{2}k_BT$, the value suggested by the equipartition of energy.  To check this, we need only note that in the high temperature limit
\begin{equation}
\label{Eq45i}
{\rm coth}(\beta\hbar\omega/2) \rightarrow \frac{2}{\beta\hbar\omega} = \frac{2k_BT}{\hbar\omega} \, .
\end{equation}
It follows that the high temperature limits of the kinetic and potential energies are, respectively:
\begin{eqnarray}
\label{Eq45j}
\left\langle \frac{\hat{p}^2}{2m} \right\rangle_{T\rightarrow\infty} &=& \frac{k_BT}{2}  \nonumber \\
\left\langle \frac{m f_0^2\hat{x}^2}{2} \right\rangle_{T\rightarrow\infty}   &=& \frac{k_BT}{2}  f_0^2 \langle\langle \omega^{-2}\rangle\rangle \, .
\end{eqnarray}
The kinetic energy of the oscillator tends to the expected high-temperature value, but the potential energy does not, and requires an explanation.  In pursuit of this, we note that the Cauchy-Schwartz inequality requires that $\langle\langle \omega^2\rangle\rangle\langle\langle \omega^{-2}\rangle\rangle \geq 1$ and, as $\langle\langle \omega^2\rangle\rangle = \Omega_0^2$, it follows that $\Omega_0^2 \langle\langle \omega^{-2}\rangle\rangle \geq 1$, so that the potential energy, when expressed in terms of $\Omega_0$, exceeds that assigned by equipartition.  The natural way to understand this is that the oscillator is strongly rather than weakly coupled to its environment and the excess thermal energy has its origin in the interaction energy with the environment.  The issue is less clear, however, if we express the potential energy in terms of $\omega_0$.  We shall see below that in the limit of weak damping, when $\Omega_0$ and $\omega_0$ tend to a common value, the probability distribution $\pi(\omega)$ becomes sharply peaked around $\omega = \Omega_0$ so that the equipartition of energy for the potential energy is restored.


\section{Oscillator dynamics}

It remains to consider the evolution of the oscillator towards equilibrium.  This will be important in practical applications of the theory but also for a fundamental reason; we have obtained equilibrium states at zero and at finite temperature, but have not as yet proven that the dynamics of the oscillator causes it to evolve towards this state.  Establishing this, without approximation, is a principal aim of this section.

The exact diagonalisation of the Hamiltonian makes it straightforward to evaluate the time-evolution of any desired property of the oscillator.  All that we need do is to express the desired observable in terms of the eigenoperators, $\hat{B}(\omega)$ and $\hat{B}^\dagger(\omega)$ and then use the time evolution of these operators, the form of which is an elementary consequence of the fact that they are energy eigenoperators:
\begin{eqnarray}
\label{Eq46}
\hat{B}(\omega,t) = \hat{B}(\omega,0)e^{-i\omega t} \nonumber \\
\hat{B^\dagger}(\omega,t) = \hat{B}^\dagger(\omega,0)e^{i\omega t} .
\end{eqnarray}
In particular, we can determine the time-evolution of the annihilation operator for the oscillator in this way:
\begin{eqnarray}
\label{Eq47}
\hat{a}(t) &=& \int_0^\infty d\omega \left[\alpha^*(\omega)\hat{B}(\omega,0)e^{-i\omega t}
-\beta(\omega)\hat{B}^\dagger(\omega,0)e^{i\omega t}\right]  \nonumber \\
&=& \left.  \int_0^\infty d\omega \right\{\alpha^*(\omega)e^{-i\omega t}
\left[\alpha(\omega)\hat{a}(0) + \beta(\omega)\hat{a}^\dagger(0) +  \right. \nonumber \\
& & \qquad  \int_0^\infty d\omega' \left. \: \left(\gamma(\omega,\omega')\hat{b}(\omega',0) + 
\delta(\omega,\omega')\hat{b}^\dagger(\omega',0)\right) \right] \nonumber \\
& & \qquad -\beta(\omega)e^{i\omega t}\left[\alpha^*(\omega)\hat{a}^\dagger(0) + \beta^*(\omega)\hat{a}(0) + \right. 
\nonumber \\
& & \qquad  \left. \int_0^\infty d\omega' \left.\: \left(\gamma^*(\omega,\omega')\hat{b}^\dagger(\omega',0) + 
\delta^*(\omega,\omega')\hat{b}(\omega',0)\right) \right] \right\} .
\end{eqnarray}
This may be used, together with the initial state of the oscillator and the environment, to evaluate the expectation value of any desired property.  Note that we have chosen the initial state to be one in which the oscillator and the reservoir are uncorrelated.  We allow the oscillator to be prepared in any chosen state, but the reservoir is in a thermal state, which we describe using a thermal vacuum state for the reservoir degrees of freedom, as given in Eq (\ref{EqC11}).  As the environment is in a stationary state, so that $\langle\hat{b}(\omega,0)\rangle = 0 = \langle\hat{b}^\dagger(\omega,0)\rangle $, the expectation values of the position and momentum operators take a pleasingly simple form:
\begin{eqnarray}
\label{Eq48}
\langle\hat{x}(t)\rangle = \langle\langle\cos(\omega t)\rangle\rangle \langle\hat{x}(0)\rangle 
+ \frac{1}{m}\langle\langle\omega^{-1}\sin(\omega t)\rangle\rangle \langle\hat{p}(0)\rangle \nonumber \\
\langle\hat{p}(t)\rangle = \langle\langle\cos(\omega t)\rangle\rangle \langle\hat{p}(0)\rangle 
- m\langle\langle\omega\sin(\omega t)\rangle\rangle \langle\hat{x}(0)\rangle ,
\end{eqnarray}
where the double angle brackets denote, as before, averages over our probability distribution $\pi(\omega)$ as in (\ref{Eq19}).  The generality of this evolution follows simply from the linearity of the dynamics and has been noted before, in particular by Haake and Reibold in their treatment of an oscillator coupled to a quasi-continuum of oscillators \cite{Haake}. The form of these equations adds further support to our interpretation of $\pi(\omega)$ as a frequency probability distribution for the damped oscillator, as they may be viewed as the evolution of an undamped oscillator with a frequency $\omega$ averaged using this probability distribution.  The dissipation arises simply from a dephasing amongst the different frequency components.

The evolution of the mean position and momentum, as given in (\ref{Eq48}), has the necessary short-time form of that for an undamped oscillator
\begin{eqnarray}
\label{Eq49}
\langle\hat{x}(\delta t)\rangle = \langle\hat{x}(0)\rangle + \frac{\langle\hat{p}(0)\rangle}{m}\delta t \nonumber \\
\langle\hat{p}(\delta t)\rangle = \langle\hat{p}(0)\rangle - m\Omega_0^2\langle\hat{x}(0)\rangle \delta t ,
\end{eqnarray}
where we have used the identity $\langle\langle\omega^2\rangle\rangle = \Omega_0^2$.  The effects of the coupling to the environment enter at order $\delta t^3$ and this is an indication of the essentially non-Markovian nature of the strongly-damped oscillator.  Our primary interest is in strongly damped oscillators and so we should note that (\ref{Eq48}) includes the possibilities of both critically-damped and over-damped evolution.  The equations contain, moreover, a simple criterion for these, which we may express in terms of our probability distribution.  The motion will be oscillatory if $\langle\langle\cos(\omega t)\rangle\rangle$ has stationary points at times other than at $t=0$.  Alternatively, we may state that the motion is under-damped if the derivative of this quantity, that is $\langle\langle\omega\sin(\omega t)\rangle\rangle$, is zero for any time other than $t=0$.  If it is zero only at $t=0$ then the motion is critically-damped or over-damped.

\subsection{Steady state}

Our expression for the evolved annihilation operator (\ref{Eq47}), together with the corresponding one for the creation operator provide a full description of the oscillator dynamics.  This is true for any initial state of the oscillator and, moreover, for any environmental state including, of course, that associated with a finite temperature.  As an illustration let us examine the evolution of the characteristic function foran arbitrary initial state of the oscillator coupled to a finite-temperature environment at time $t = 0$. The zero-temperature behaviour follows, simply, in the limit $T \rightarrow 0$ or $\beta \rightarrow \infty$. With a little effort we find (using the method of characteristics \cite{Radmore1997})
\begin{equation}
\label{Eq50}
\chi_T[\xi,t] = \chi_T[\xi(t),0]\exp\left[-\frac{1}{2}\int_0^\infty d\omega' 
\left|\int_0^\infty d\omega \: \mu(\omega,\omega',t)\right|^2\right] ,
\end{equation}
where
\begin{eqnarray}
\label{Eq51}
\xi(t) = \int_0^\infty d\omega \left[\eta(\omega,t)\alpha^*(\omega) - \eta^*(\omega,t)\beta(\omega)\right]
\nonumber \\
\mu(\omega,\omega',t) = \eta(\omega,t)\gamma^*(\omega,\omega') - \eta^*(\omega,t)\delta^*(\omega,\omega')
\nonumber \\
\eta(\omega,t) = \left[\xi\alpha(\omega)+\xi^*\beta(\omega)\right]{\rm coth}(\beta\hbar\omega/2)e^{i\omega t} .
\end{eqnarray}
This characteristic function encodes the full dynamics and statistics of the oscillator.  As a simple illustration of this we can determine, directly, the form of the steady state.  To see this we first note $\xi(t)$ tends to zero as $t$ tends to infinity and the different frequency components dephase so that the prefactor in Eq (\ref{Eq50}), corresponding to the initial state of the oscillator tends to
\begin{equation}
\label{Eq52}
\chi[\xi(\infty),0] = \chi[0,0] = 1 ,
\end{equation}
which means that all memory of the initial state of the oscillator is lost. Evaluating the long-time limit of the exponential factor in (\ref{Eq50}) requires some care in the handling of the delta-function and principal part components.  We find 
\begin{equation}
\label{Eq53}
\int_0^\infty d\omega' 
\left|\int_0^\infty d\omega \: \mu(\omega,\omega',\infty)\right|^2
 = \int_0^\infty d\omega \left|\xi\alpha(\omega') + \xi^*\beta(\omega')\right|^2 {\rm coth}(\beta\hbar\omega/2) ,
\end{equation}
so the steady-state characteristic function is
\begin{equation}
\label{Eq54}
\chi(\xi,\infty) = \exp\left(-\frac{1}{2}\int_0^\infty d\omega\left|\xi\alpha(\omega) + \xi^*\beta(\omega)\right|^2{\rm coth}(\beta\hbar\omega/2) \right) ,
\end{equation}
which we recognise as the characteristic function for the oscillator in the global thermal equilibrium state (\ref{Eq45e}). This is a most satisfactory and exact result.

It also means that the steady state of the oscillator is the mean-force Gibbs state.  To see this we need only note that it is given by the trace over the environment of the full thermal equilibrium state, Eq (\ref{Eq45b}). Proof of this equivalence has also been shown in \cite{Subasi2012} by demonstrating the equality of steady state multi-time open system correlation functions obtained by Heisenberg-Langevin equation of motion methods to those of the closed system thermal Gibbs state. 

\subsection{An example evolution}

We have seen that both the dynamics and the steady-state of our damped harmonic oscillator are governed by the form of the function $\pi(\omega)$.  Determining this, together with the initial conditions, provides all the information required.  We can calculate $\pi(\omega)$ directly from the frequency-dependence of the coupling between the oscillator and its environment or, more simply, select a form for $\pi(\omega)$ and proceed from this.  This is the approach we adopt here.

As an example we consider a $\pi(\omega)$ of the form
\begin{equation}
\label{Ex1}
\pi(\omega) = \frac{2\omega^2}{\pi} \cdot \frac{(\gamma_+ + \gamma_-)(\gamma_- + \Gamma)(\Gamma + \gamma)}
{(\omega^2 + \Gamma^2)(\omega^2 + \gamma_+^2)(\omega^2 + \gamma_-^2)} \, .
\end{equation}
This is, perhaps, the simplest example that satisfies the necessary physical constraints: (i) it is normalised, (ii) $\pi(0) = 0$, and (iii) $ \langle \langle \omega^2 \rangle \rangle$ is finite.  We would like our example to lead to an evolution that is close to the familiar classical behaviour. To this end we choose $\Gamma$ to be a real decay rate and $\gamma_\pm$ are either two real decay rates or complex conjugates of one another.  With the familiar classical behaviour in mind we set
\begin{equation}
\label{Ex2}
\gamma_\pm = \frac{\gamma}{2} \pm \sqrt{\frac{\gamma^2}{4} - \omega_0^2} .
\end{equation}
Direct evaluation of $\langle \langle \omega^2 \rangle \rangle$ for our example gives
\begin{equation}
\label{Ex3}
\langle\langle\omega^2\rangle\rangle = \Omega_0^2 = \Gamma(\gamma_+ + \gamma_-) + \gamma_+\gamma_- \, ,
\end{equation}
which links the parameters $\Gamma$, $\gamma_+$ and $\gamma_-$ to the short-time natural frequency $\Omega_0$.  When written in terms of $\omega_0$ and $\gamma$, this relationship simplifies to
\begin{equation}
\label{Ex4}
\Omega_0^2 = \omega_0^2 + \Gamma\gamma \, .
\end{equation}
We can rewrite this expression in the form
\begin{equation}
\label{Ex5}
\Gamma = \frac{\Omega_0^2 - \omega_0^2}{\gamma} \, ,
\end{equation}
which emphasises the need for the decay rate $\Gamma$ as a consequence of the fact that we require two natural frequencies, $\omega_0$ and $\Omega_0$, to describe the quantum damped harmonic oscillator.  It is straightforward to determine from $\pi(\omega)$ the two averages $\langle\langle\omega\rangle\rangle$ and $\langle\langle\omega^{-1}\rangle\rangle$:
\begin{eqnarray}
\label{Ex6}
\langle\langle\omega\rangle\rangle &=&  \frac{2}{\pi} \cdot 
\frac{\gamma_+^2\gamma_-^2\ln(\gamma_-/\gamma_+) + \gamma_-^2\Gamma^2\ln(\Gamma/\gamma_-) + \Gamma^2\gamma_+^2\ln(\gamma_+/\Gamma)}
{(\Gamma - \gamma_+)(\gamma_+ - \gamma_-)(\gamma_- - \Gamma)}
\nonumber \\
\langle\langle\omega^{-1}\rangle\rangle &=&  \frac{2}{\pi} \cdot 
\frac{\Gamma^2\ln(\gamma_-/\gamma_+) + \gamma_+^2\ln(\Gamma/\gamma_-) + \gamma_-^2\ln(\gamma_+/\Gamma)}
{(\Gamma - \gamma_+)(\gamma_+ - \gamma_-)(\gamma_- - \Gamma)}
\end{eqnarray}
from which we can derive the expectation values of $\hat{x}^2$ and $\hat{p}^2$ in the ground state of the damped oscillator \footnote{When $\gamma_\pm$ are complex, we can write these in the 
form $\gamma_\pm = |\gamma_+|e^{\pm i \phi}$, where $\phi = \tan^{-1}[\Im(\gamma_+)/\Re(\gamma_+)]$.}.

We can determine the evolution of the expectation values of the position and momentum from Eq (\ref{Eq48}) by evaluating the averages $\langle\langle\cos(\omega t)\rangle\rangle$, $\langle\langle\omega^{-1}\sin(\omega t)\rangle\rangle$ and $\langle\langle\omega\sin(\omega t)\rangle\rangle$:
\begin{eqnarray}
\label{Ex7}
\langle\langle\cos(\omega t)\rangle\rangle &=& \frac{\Gamma(\gamma_+ + \gamma_-)}{(\Gamma - \gamma_+)(\gamma_- -\Gamma)}e^{-\Gamma t}
 + \frac{\gamma_+(\Gamma + \gamma_-)}{(\Gamma - \gamma_+)(\gamma_+ - \gamma_-)}e^{-\gamma_+ t}  \nonumber \\
& &   \qquad \qquad + \frac{\gamma_-(\Gamma + \gamma_+)}{(\gamma_- - \Gamma)(\gamma_+ - \gamma_-)}e^{-\gamma_- t} \nonumber  \\
\langle\langle\omega^{-1}\sin(\omega t)\rangle\rangle &=& \int_0^t \langle\langle\cos(\omega t')\rangle\rangle dt' \nonumber  \\
\langle\langle\omega\sin(\omega t)\rangle\rangle &=& -\frac{d}{dt}\langle\langle\cos(\omega t)\rangle\rangle    \,  .
\end{eqnarray}
We expect to find something approximating the evolution of the classical damped harmonic oscillator in the limit of small $\Gamma$, but departures from this for larger values.  In figure 2 we plot the evolution of $\langle\langle\cos(\omega t)\rangle\rangle$ for a small value of $\Gamma$.  As anticipated, we find over-damped like behaviour for real values of $\gamma_+$ and $\gamma_-$, and underdamped behaviour for complex values.  Note, however, that there is a small overshoot in the underdamped regime, which would certainly be absent in the over-damped evolution of the classical oscillator.  This can be traced back to the $e^{-\Gamma t}$ term which, although small, decays slowly and so has a residual influence at long times. This is especially clear in the large $\Gamma$ regime , depicted in figure 3.  There we see that there is a significant over-shoot of the mean position in what, classically, would be the over-damped regime. In the under-damped regime, however, the $e^{-\Gamma t}$ term has a less dramatic effect; the evolution for $\Gamma = 0.01$ (in figure 2b) and for $\Gamma = 10$ (figure 3b) are qualitatively rather similar.

\begin{figure}[htbp] 
\centering
\includegraphics[width=10cm]{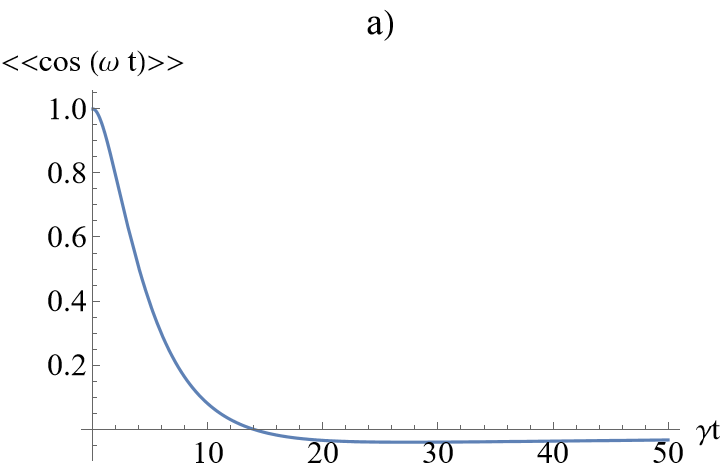}
\includegraphics[width=10cm]{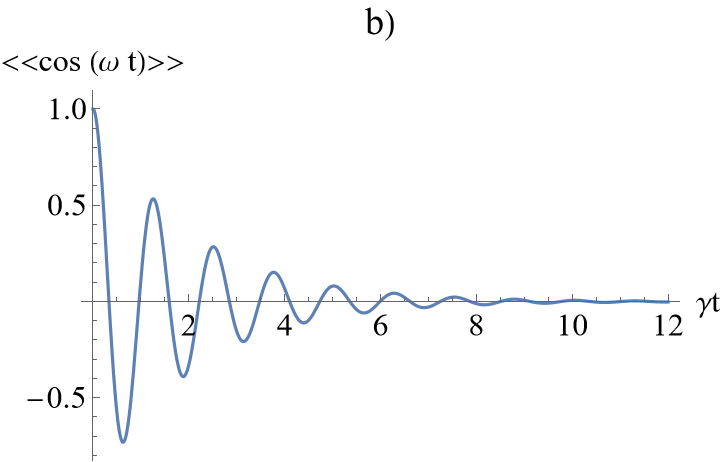}
\caption{Evolution of $\langle \langle \cos (\omega t) \rangle \rangle$ for small $\Gamma$. a) Overdamped regime, with parameters $\Gamma = 0.01 \gamma$, $\gamma_+ = \frac{3}{4} \gamma$, $\gamma_-=\frac{1}{4} \gamma$; b) Underdamped regime with parameters $\Gamma = 0.01 \gamma$, $\gamma_\pm = \left( \frac{1}{2} \pm 5 i \right) \gamma$.} 
\label{fig:figure2}
\end{figure}

\begin{figure}[htbp] 
\centering
\includegraphics[width=10cm]{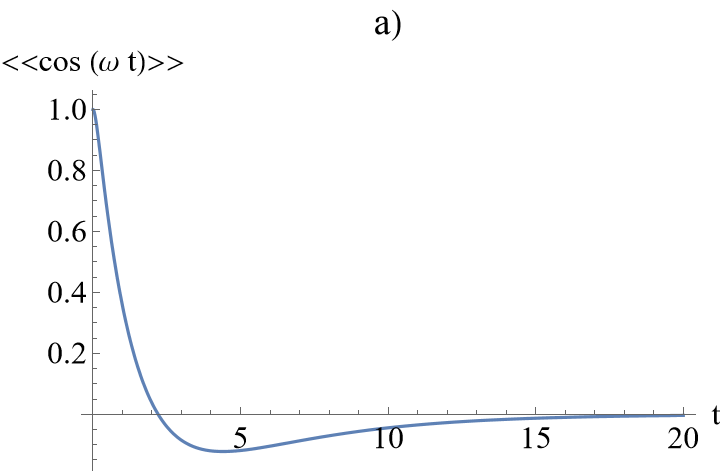}
\includegraphics[width=10cm]{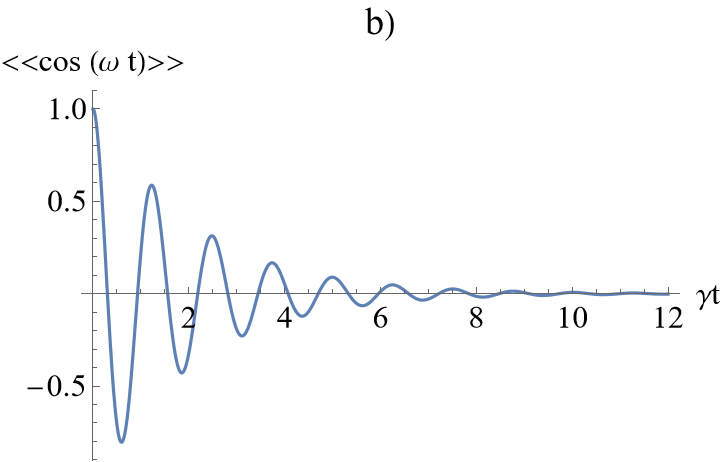}
\caption{Evolution of $\langle \langle \cos (\omega t) \rangle \rangle$ for large $\Gamma$. a) Overdamped regime, with parameters $\Gamma = 10 \gamma$, $\gamma_+ = \frac{3}{4} \gamma$, $\gamma_-=\frac{1}{4} \gamma$; b) Underdamped regime with parameters $\Gamma = 10 \gamma$, $\gamma_\pm = \left( \frac{1}{2} \pm 5 i \right) \gamma$.} 
\label{fig:figure3}
\end{figure}

We have seen that the existence of two natural frequencies for the oscillator, $\omega_0$ and $\Omega_0$, is particularly important when comparing the short and long time behaviour of the oscillator: $\omega_0$ behaves as the natural frequency, but at short times it is $\Omega_0$ that takes on this role.  In figure 4 we plot the short-time evolution of $\langle\langle\cos(\omega t)\rangle\rangle$ (solid line) and compare this with the classical evolution of for a classical damped harmonic oscillator with natural frequency $\omega_0$  It is clear that the former falls off more quickly as it must, because $\Omega_0 > \omega_0$. For the parameters chosen, $\Omega_0$ is just over twice the value of $\omega_0$. It is interesting to note that we can match the two evolutions up to very short times if we allow for a short-time slip so that as $t$ tends to zero, $\langle \langle \cos (\omega t) \rangle \rangle$ becomes larger than zero \cite{Haake}. This is unnecessary, however, if we take account of the existence of two natural frequencies as we have done here.

\begin{figure}[htbp] 
\centering
\includegraphics[width=10cm]{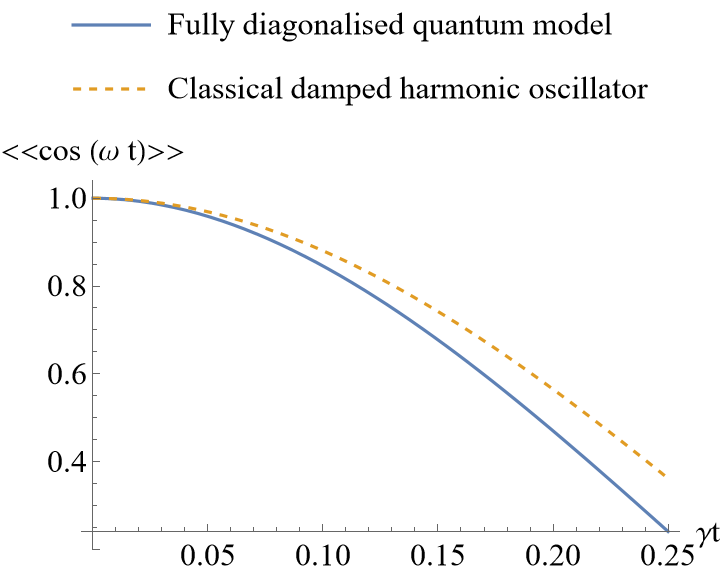}
\caption{Comparison of short-time behaviour in quantum and classical cases. The exact solution for short times is plotted as a solid line, with $\pi(\omega)$ as given in Eq.~\ref{Ex1}, and parameters $\Gamma = 10 \gamma$, $\gamma_+ = \frac{3}{4} \gamma$, $\gamma_-=\frac{1}{4} \gamma$; the dashed line shows the limit $\Gamma \rightarrow 0$, in which classical damped simple harmonic motion is recovered, with the other parameters unchanged, $\gamma_\pm = \left( \frac{1}{2} \pm 5 i \right) \gamma$} 
\label{fig:figure4}
\end{figure}

\section{Weak-coupling limit}
The theory developed above was designed to treat the strongly-damped harmonic oscillator, but should also be applicable to the more familiar weakly damped oscillator, for which the oft-employed Born and Markov approximations are applicable and the steady state of the oscillator should be its ground state. We show here that this is indeed the case.

We start by considering the form of the function $\alpha(\omega)$ in the weak coupling limit.  To aid our analysis we rewrite the form given in (\ref{EqA18}) as
\begin{equation}
\label{Eq55}
\alpha(\omega) = \frac{\omega + \Omega_0}{\Omega_0}\left(\frac{V(\omega)}{|V(\omega)|^2Y(\omega)
-i\pi|V(\omega)|^2}\right) .
\end{equation}
The weak damping limit corresponds to choosing the coupling to the environment to be small or, more specifically, to $|V(\omega)|^2 \ll \Omega_0$.  It is clear that in this limit, $|\alpha(\omega)|^2$ will be a sharply peaked function centred around the frequency for which $Y(\omega) = 0$.  If the integral part in $Y(\omega)$, as given in (\ref{EqA15}) is small\footnote{If it is not then we will need to invoke ideas of renormalisation, a manageable complication, but one we wish to avoid.} then this frequency will be close to $\Omega_0$ and we can write
\begin{equation}
\label{Eq56}
|V(\omega)|^2Y(\omega) \approx 4(\omega - \Omega_0) - 4F(\omega) ,
\end{equation}
where 
\begin{equation}
\label{Eq57}
4F(\omega) = \int_0^\infty d\omega' 
\left(\frac{\mathbbmss{P}}{\omega-\omega'} - \frac{1}{\omega+\omega'}\right)|V(\omega')|^2 .
\end{equation}
This leads, in turn, to a corresponding approximate form for $\alpha(\omega)$:
\begin{equation}
\label{Eq58}
\alpha(\omega) \approx \frac{V(\omega)}{2}\frac{1}{\omega - \Omega_0 - F(\omega) - i\frac{\pi}{4}|V(\omega)|^2} ,
\end{equation}
where we have set $\omega = \Omega_0$ everywhere except in the denominator.  We note that this is of the form that arises from the Fano diagonalisation of our problem if we make the rotating wave approximation by omitting from our original Hamiltonian all terms that are products of two creation operators or of two annihilation operators \cite{Radmore1997,Radmore1988}.

Consistency with the above approximation, which led us to set $\omega = \Omega_0$ leads us to set $\beta(\omega)$ to zero:
\begin{equation}
\label{Eq59}
\beta(\omega) \approx 0 ,
\end{equation}
so that the integral over all frequency of $|\alpha(\omega)|^2$ is unity.  Moreover, for weak damping the thermal function ${\rm coth}(\beta\hbar\omega/2)$ will also be slowly varying compared to the rapid variation of $|\alpha(\omega)|^2$ in the vicinity of $\omega = \Omega_0$ and we may replace this function by its value at $\Omega_0$.  Hence in this limit the steady-state characteristic function for our oscillator is
\begin{equation}
\label{Eq71}
\chi(\xi,\infty) = \exp\left(-\frac{1}{2}|\xi|^2{\rm coth}(\beta\hbar\Omega_0/2)\right) 
= \exp\left(-\frac{1}{2}|\xi|^2(2 \bar{n} + 1)\right)  ,
\end{equation}
where $\bar{n}$ is the mean thermal excitation number. We recognise (\ref{Eq71}) as the symmetrically ordered characteristic function for the thermal state of the undamped oscillator \cite{Radmore1997}, as it must be. Further, we note that in this limit our probability distribution function, $\pi(\omega) \approx |\alpha(\omega)|^2$, approaches a Lorentzian centred on $\Omega_0$, with some width $\gamma$ \footnote{This is not strictly true in the wings of the distribution, of course, as even in this limit we require $\langle\langle\omega^2\rangle\rangle = \Omega_0^2$, but the corresponding quantity for a true Lorentzian is divergent.}.  Thus all the complexity of of the original problem is reduced, in the weak-damping limit to just three parameters: a natural oscillation frequency, a damping rate and a temperature. 

It was important to confirm that our more general treatment coincided, in the right limit, with the approximate methods used for weakly damped oscillators.  We should note that even if we are working in the weakly damped regime, then our approach offers a systematic way to treat corrections to the results obtained using the Born and Markov approximations, which may play an important role in modelling measurements at the limits of sensitivity.


\section{Conclusion}

We have presented an exact diagonalisation of a simple quantum model of the damped harmonic oscillator, one that is applicable, in particular, to  any strength of the damping.  As a result we have recovered the fact that much of the behaviour of the oscillator and many of its properties can be described in terms of a single probability function, $\pi(\omega)$, which we may interpret as the contribution of corresponding dressed mode, at frequency $\omega$ to the oscillator.  These properties include the steady state of the oscillator at both at zero and at finite temperature, the entanglement between the oscillator and its environment and also its evolution, both in the familiar under-damped regime but also in the more problematic over-damped regime.  

We have applied our diagonalisation to study the properties of the true ground state and have shown that the oscillator part of this pure entangled state coincides with the steady-state of the oscillator in a zero-temperature environment.  The diagonalisation is not specific to any particular state of the reservoir, however, and we have shown how it can be be applied to environments at finite temperature. The extension to more exotic states, such as squeezed reservoirs presents no obvious difficulties.  It may be extended, moreover, to include driving forces, coupled oscillators and multiple reservoirs, with the latter perhaps being at different temperatures \cite{Martinez}.  This may provide some insights into important questions of principle in the nascent fields of quantum machines and quantum thermodynamics \cite{Gemmer, Vinjanampathy2016,Joan2011,Joan2013,Roncaglia,Binder,Brandao}.  


\ack
It is a pleasure to dedicate this paper to our friend and colleague, Igor Jex, in celebration of his 60th birthday.

The Hamiltonian diagonalisation upon which much of this work is based was first calculated by Bruno Huttner, in collaboration with SMB, 30 years ago in their work on the quantum electrodynamics of dielectric media.  We are most grateful to him and also to Paul Radmore and Claire Gilson for helpful comments and suggestions.  This work was supported, in part, by the Royal Society through the award to SMB of a Research Professorship, RP150122.


\appendix

\section{Heisenberg equations of motion}
\label{AppHeisenberg}

The Heisenberg equations of motion follow directly from the Hamiltonian, and we obtain these using the Hamiltonian in the form of (\ref{Eq2}).  (We could equally well have used the identical Hamiltonian (\ref{Eq3}).)
\begin{eqnarray}
\label{EqH.1}
\dot{\hat{x}} &=& \frac{\hat{p}}{m}  \nonumber \\
\dot{\hat{p}} &=& -m\Omega_0^2\hat{x} + \sum_\mu m_\mu \omega_\mu^2 \hat{x}_\mu  \nonumber \\
\dot{\hat{x}}_\mu &=& \frac{\hat{p}_\mu}{m_\mu}   \nonumber \\
\dot{\hat{p}}_\mu &=& -m_\mu\omega_\mu^2 \hat{x}_\mu + m_\mu\omega^2_\mu \lambda_\mu \hat{x} \, .
\end{eqnarray}
We seek an equation of motion for the position operator and so first eliminate the momentum operators between the first and second and the third and fourth equations:
\begin{eqnarray}
\label{EqH.2}
\ddot{\hat{x}} + \omega_0^2 \hat{x} &=& \sum_\mu \frac{m_\mu}{m}\omega^2_\mu\lambda_\mu \hat{x}_\mu \\
\label{EqH.3}
\ddot{\hat{x}}_\mu + \omega_\mu^2 \hat{x}_\mu &=& \omega^2_\mu \lambda_\mu \hat{x}  \, .
\end{eqnarray}
The next step is to integrate the second of these equations of motion.  The complementary function is
\begin{eqnarray}
\label{EqH.4}
\hat{x}_\mu^{\rm CF} &=& \hat{x}_\mu(0) \cos(\omega_\mu t) + \frac{\dot{\hat{x}}_\mu(0)}{\omega_\mu}\sin(\omega_\mu t)
\nonumber \\
&=& \hat{x}_\mu(0) \cos(\omega_\mu t) + \frac{\hat{p}_\mu(0)}{m_\mu\omega_\mu}\sin(\omega_\mu t) \, .
\end{eqnarray}
To find the particular integral we need to make a small addition to (\ref{EqH.4}) by adding a very weak damping term to give:
\begin{equation}
\label{EqH.5}
\ddot{\hat{x}}_\mu + \varepsilon\dot{\hat{x}}_\mu + \omega_\mu^2 \hat{x}_\mu = \omega^2_\mu \lambda_\mu \hat{x}  \, ,
\end{equation}
and work in the limit as the strictly positive quantity $\varepsilon $ tends to zero.  It is this choice of a positive (if very small) value for $\varepsilon$ that provides the irreversibility and hence the arrow of time.

Solving (\ref{EqH.5}) for the particular integral requires some care and so we pause for a moment to provide the details. Let us introduce the Fourier transform of $\hat{x}_\mu$ in the form
\begin{equation}
\label{EqH.6}
\bar{x}_\mu(\omega) = \frac{1}{\sqrt{2\pi}}\int_{-\infty}^\infty \hat{x}_\mu(t) e^{i\omega t}dt \, ,
\end{equation}
with a similar expression for the Fourier transform of $\hat{x}$.  It follows that the Fourier transform of the particular integral part of the position is given by
\begin{eqnarray}
\label{EqH.7}
-\omega^2 \bar{x}_\mu -i\varepsilon\bar{x}_\mu + \omega_\mu^2\bar{x}_\mu &=& \omega_\mu^2\lambda_\mu \bar{x}
\nonumber \\
\qquad \qquad \Rightarrow  \qquad\bar{x}_\mu &=& -\frac{\omega_\mu^2\lambda_\mu \bar{x}}{\omega^2 -\omega_\mu^2 + i\varepsilon\omega} \, .
\end{eqnarray}
It follows then follows that
\begin{eqnarray}
\label{EqH.8}
\hat{x}_\mu^{\rm PI}(t)  &=&  -\frac{1}{\sqrt{2\pi}}\int_{-\infty}^\infty \frac{\omega_\mu^2\lambda_\mu}{\omega^2 -\omega_\mu^2 + i\varepsilon\omega}
\bar{x}(\omega) e^{-i\omega t}d\omega \, ,
\end{eqnarray}
which is the Fourier transform of the product of two functions of $\omega$.  We can use the convolution theorem to write this in terms of the transforms.  To exploit this we write the Fourier transform of the first function as the time-derivative of a function $K(t)$:
\begin{eqnarray}
\label{EqH.9}
\dot{K}_\mu(t) &=& \frac{1}{2\pi}\int_{-\infty}^\infty \frac{\omega_\mu^2\lambda_\mu}{\omega^2 -\omega_\mu^2 + i\varepsilon\omega}
e^{-i\omega t}d\omega  \nonumber \\
&=& -\omega_\mu\lambda_\mu e^{-\varepsilon t/2}\sin(\omega_\mu t) 
\end{eqnarray}
if $t  > 0$ and is zero otherwise.  We can now take the limit as $\varepsilon \rightarrow 0$ to give
\begin{eqnarray}
\label{EqH.10}
\dot{K}_\mu &=& -\omega_\mu\lambda_\mu \sin(\omega_\mu t)  \nonumber \\
\Rightarrow K_\mu &=& \lambda_\mu \cos(\omega_\mu t) \, .
\end{eqnarray}
It then follows that 
\begin{eqnarray}
\label{EqH.11}
\hat{x}_\mu^{\rm PI}(t)  &=& -\int_{-\infty}^\infty \bar{\dot{K}}_\mu(\omega)\bar{x}(\omega) e^{-i\omega t}d\omega \nonumber \\
&=& -\int_{-\infty}^\infty \dot{K}_\mu(T) \hat{x}(t-T)dT \nonumber \\
&=& -\left[K(T)\hat{x}(t-T)\right]_0^t + \int_0^t K_\mu(T)\frac{d}{dT}\hat{x}(t-T) dT \nonumber \\
&=& -K_\mu(t)\hat{x}(0) + K_\mu(0)\hat{x}(t)  - \int_0^t K_\mu(t-t') \dot{\hat{x}}(t') dt' \, ,
\end{eqnarray}
where we have used the facts that $K(T) = 0$ for $T < 0$ and we may take $\hat{x}(\tau) = 0$ for $\tau < 0$.

Pulling this altogether we arrive at our desired Heisenberg-Langevin equation for the damped harmonic oscillator. From (\ref{EqH.3}) we have
\begin{equation}
\label{EqH.12}
\ddot{\hat{x}}(t) + \int_0^t \kappa(t-t') \dot{\hat{x}}(t') dt' + \left(\Omega_0^2 - \kappa(0)\right) \hat{x}(t) 
+ \kappa(t)\hat{x}(0) = \frac{\hat{F}(t)}{m} \, ,
\end{equation}
where 
\begin{eqnarray}
\label{EqH.13}
\kappa(t) &=& \sum_\mu \frac{m_\mu}{m}\Omega_\mu^2\lambda_\mu K(t)  \nonumber \\
&=& \sum_\mu \frac{m_\mu}{m}\omega_\mu^2\lambda_\mu^2 \cos(\omega_\mu t)
\end{eqnarray}
and $F(t)$ is the Langevin force:
\begin{equation}
\label{EqH.14}
\hat{F}(t) = \omega^2_\mu\lambda_\mu\left(\hat{x}_\mu(0) \cos(\omega_\mu t) + \frac{\hat{p}_\mu(0)}{m_\mu\omega_\mu}\sin(\omega_\mu t)\right) \, .
\end{equation}

\section{Fano diagonalisation}
\label{Fano}

We present a brief account of the exact diagonalisation of our Hamiltonian based on methods developed by Fano for the study of configuration interactions \cite{Fano}.  This idea was applied extended to weakly-coupled oscillators in a quantum study of damped cavity modes \cite{Radmore1997,Radmore1988}.  The extension to stronger couplings, with the inclusion of counter-rotating couplings, has been given before and applied to the quantum theory of light in dielectric media \cite{Huttner1992a,Huttner1992b}.  We summarise here the analysis presented in \cite{Huttner1992b}.

Our task is to diagonalise the damped harmonic oscillator Hamiltonian\footnote{For the Hamiltonian of interest in this paper the coupling, $V(\omega)$, is real but treating the problem with a more general complex coupling presents no additional difficulties.}
\begin{eqnarray}
\label{EqA1}
\hat{H} = \hbar\Omega_0\hat{a}^\dagger\hat{a} + \int_0^\infty d\omega \:\hbar\omega \hat{b}^\dagger(\omega)\hat{b}(\omega) \nonumber \\
  \qquad \qquad +\int_0^\infty d\omega \: \frac{\hbar}{2}\left(\hat{a} + \hat{a}^\dagger\right)\left[V(\omega)\hat{b}^\dagger(\omega)
+V^*(\omega)\hat{b}(\omega)\right] ,
\end{eqnarray}
by which we mean rewriting it in the form of a continuum of \emph{uncoupled} or dressed oscillators:
\begin{equation}
\label{EqA2}
\hat{H} = \int_0^\infty d\omega \:\hbar\omega\: \hat{B}^\dagger(\omega)\hat{B}(\omega) + C  ,
\end{equation}
where $C$ is an unimportant constant.  We proceed by writing the dressed annihilation operators, $\hat{B}(\omega)$, as linear combinations of the bare operators for the oscillator and bath modes:
\begin{equation}
\label{EqA3}
\hat{B}(\omega) = \alpha(\omega)\hat{a} + \beta(\omega)\hat{a}^\dagger + 
\int_0^\infty d\omega' \: \left[\gamma(\omega,\omega')\hat{b}(\omega') + \delta(\omega,\omega')\hat{b}^\dagger(\omega')\right] ,
\end{equation}
where $\alpha{\omega}$, $\beta(\omega)$, $\gamma(\omega,\omega')$ and $\delta(\omega,\omega')$ are to be determined.

We require the operator $\hat{B}(\omega)$ to be associated with an uncoupled or dressed oscillator of angular frequency $\omega$.  This requires us to find its form such that the following pair of operator equations are satisfied for every frequency, $\omega$:
\begin{eqnarray}
\label{EqA4}
\left[\hat{B}(\omega), \hat{H}\right] = \hbar\omega \hat{B}(\omega) \\
\label{EqA5}
\left[\hat{B}(\omega),\hat{B}^\dagger(\omega')\right] = \delta(\omega-\omega')  .
\end{eqnarray}
Substituting the ansatz (\ref{EqA3}) into (\ref{EqA4}) and comparing coefficients of the bare creation and annihilation operators leads to the set of coupled equations:
\begin{eqnarray}
\label{EqA6}
\alpha(\omega)\Omega_0 + \frac{1}{2}\int_0^\infty d\omega' \left[\gamma(\omega,\omega')V(\omega')
- \delta(\omega,\omega')V^*(\omega')\right] = \alpha(\omega)\omega \\
\label{EqA7}
-\beta(\omega)\Omega_0 + \frac{1}{2}\int_0^\infty d\omega' \left[\gamma(\omega,\omega')V(\omega')
- \delta(\omega,\omega')V^*(\omega')\right] = \beta(\omega)\omega \\
\label{EqA8}
\frac{V^*(\omega')}{2}\left[\alpha(\omega) - \beta(\omega)\right] + \gamma(\omega,\omega')\omega'    = \gamma(\omega,\omega')\omega \\
\label{EqA9}
\frac{V(\omega')}{2}\left[\alpha(\omega) - \beta(\omega)\right] - \delta(\omega,\omega')\omega'    = \delta(\omega,\omega')\omega  .
\end{eqnarray}
Our method of solution is use these to determine the functions $\beta(\omega)$, $\gamma(\omega,\omega')$ and $\delta(\omega,\omega')$ in terms of $\alpha(\omega)$ and then to determine this remaining function by enforcing the commutation relation (\ref{EqA5}).  From (\ref{EqA6}) and (\ref{EqA7}) we see that
\begin{equation}
\label{EqA10}
\beta({\omega}) = \frac{\omega - \Omega_0}{\omega + \Omega_0}\: \alpha(\omega)  .
\end{equation}
If we use this to substitute for $\beta(\omega)$ into the remaining equations then we find
\begin{eqnarray}
\label{EqA11}
V^*(\omega')\frac{\Omega_0}{\omega + \Omega_0}\alpha(\omega) = \gamma(\omega,\omega')(\omega - \omega') \\
\label{EqA12}
V(\omega')\frac{\Omega_0}{\omega + \Omega_0}\alpha(\omega) = \delta(\omega,\omega')(\omega + \omega')  .
\end{eqnarray}
Solving the second of these presents no difficulty and we find
\begin{equation}
\label{EqA13}
\delta(\omega,\omega') = \left(\frac{1}{\omega + \omega'}\right)V(\omega')\frac{\Omega_0}{\omega + \Omega_0}\alpha(\omega) .
\end{equation}
The first, however, requires careful handling because the behaviour at $\omega = \omega'$.  Following Fano \cite{Fano}, we adopt the method proposed by Dirac \cite{Dirac1927} and write
\begin{equation}
\label{EqA14}
\gamma(\omega,\omega') = \left(\frac{\mathbbmss{P}}{\omega - \omega'} + Y(\omega)\delta(\omega - \omega')\right)
V^*(\omega')\frac{\Omega_0}{\omega + \Omega_0}\alpha(\omega) ,
\end{equation}
where $\mathbbmss{P}$ denotes that the principal part is to be taken on integration and $Y(\omega)$ is a real function, which we determine by substituting (\ref{EqA14}) into (\ref{EqA6}).  We find
\begin{eqnarray}
\label{EqA15}
Y(\omega) = \frac{1}{|V(\omega)|^2}\left[\frac{2(\omega^2-\Omega_0^2)}{\Omega_0} - \int_0^\infty d\omega' 
\left(\frac{\mathbbmss{P}}{\omega-\omega'} - \frac{1}{\omega+\omega'}\right)|V(\omega')|^2\right] . 
\nonumber  \\
\end{eqnarray}

If we substitute our operators, $\hat{B}(\omega)$, expressed in terms of the function $\alpha(\omega)$ into the commutation relation (\ref{EqA6}) then we find
\begin{eqnarray}
\label{EqA16}
\left[\hat{B}(\omega),\hat{B}^\dagger(\omega')\right] = \alpha({\omega})\alpha^*(\omega')\left\{1 - 
\left(\frac{\omega-\Omega_0}{\omega+\Omega_0}\right)\left(\frac{\omega'-\Omega_0}{\omega'+\Omega_0}\right)\right.
\nonumber \\
 \quad + \int_0^\infty d\omega''
\left[\left(\frac{\mathbbmss{P}}{\omega-\omega''}+Y(\omega)\delta(\omega-\omega'')\right) 
\left(\frac{\mathbbmss{P}}{\omega'-\omega''}+Y(\omega)\delta(\omega'-\omega'')\right) \right.
\nonumber \\
\qquad \qquad \left. \left. - \left(\frac{1}{\omega + \omega''}\right) \left(\frac{1}{\omega' + \omega''}\right)\right]
\frac{|V(\omega'')|^2\Omega_0^2}{(\omega+\Omega_0)(\omega'+\Omega_0)} \right\} .
\end{eqnarray}
Evaluating the integrals and setting the result equal to $\delta(\omega-\omega')$ gives\footnote{This requires the use of the following formula for the product of two principal parts \cite{Radmore1997}:
\begin{eqnarray}
\nonumber
\frac{\mathbbmss{P}}{\omega-\omega''}\frac{\mathbbmss{P}}{\omega'-\omega''} =
\frac{\mathbbmss{P}}{\omega'-\omega}\left(\frac{\mathbbmss{P}}{\omega-\omega''} -
\frac{\mathbbmss{P}}{\omega'-\omega''}\right) + \pi^2\delta(\omega-\omega'')\delta(\omega'-\omega'') .
\end{eqnarray}}
\begin{equation}
\label{EqA17}
|\alpha(\omega)|^2 = \frac{(\omega + \Omega_0)^2}{\Omega_0^2|V(\omega)|^2}
\left(\frac{1}{Y^2(\omega) + \pi^2}\right) .
\end{equation}
Note that the diagonalisation does not fix the phase of the complex function $\alpha(\omega)$ and we are free to choose this as we wish.  A convenient choice is to set
\begin{equation}
\label{EqA18}
\alpha(\omega) = \frac{\omega + \Omega_0}{\Omega_0V^*(\omega)}
\left(\frac{1}{Y(\omega) - i\pi}\right) .
\end{equation}

\section{The symmetrically ordered characteristic function}
\label{Characteristic}

We have made use of the symmetrically ordered characteristic function,
\begin{equation}
\label{EqB1}
\chi(\xi) = {\rm Tr}\left[\hat{\rho}\exp\left(\xi\hat{a}^\dagger - \xi^*\hat{a}\right)\right] \, ,
\end{equation}
to investigate both the dynamics and the steady state of the damped harmonic oscillator.  For completeness, we summarise here the main properties of this function.  Further details of this and also of related characteristic functions can be found in \cite{Radmore1997}.

The characteristic function is always defined and also well-behaved for any oscillator state.  As $\xi = 0$ it reduces to the trace of $\hat{\rho}$ and it follows that $\chi(0) = 1$.  More generally, it is the expectation value of the unitary displacement operator:
\begin{equation}
\label{EqB2}
\hat{D}(\xi) = \exp\left(\xi\hat{a}^\dagger - \xi^*\hat{a}\right) \, .
\end{equation}
This operator, by virtue of its unitarity, has only eigenvalues of modulus 1 and it follows that
\begin{equation}
\label{EqB3}
|\chi(\xi)| \leq 1 \, ,
\end{equation}
with the maximum at $\xi = 0$.

The most important property is that the density operator and the symmetrically ordered characteristic function exist in one to one correspondence, analogous to a Fourier transform pair.  Cahill and Glauber \cite{Cahill} (see also \cite{Dalton2023}) exploited a theorem of Weyl \cite{Weyl} to show that
\begin{equation}
\label{EqB2a}
\hat{\rho} = \int\frac{d^\xi}{\pi}{\rm Tr}[\hat{\rho}\hat{D}(\xi)]\hat{D}(-\xi) .
\end{equation}

We can extract from the characteristic function the expectation value of any symmetrically ordered combination of $\hat{a}$ and $\hat{a}^\dagger$:
\begin{equation}
\label{EqB4}
S\langle \hat{a}^{\dagger m}\hat{a}^n \rangle = \left. \left(\frac{\partial}{\partial\xi}\right)^m\left(-\frac{\partial}{\partial\xi^*}\right)^n
\chi(\xi) \right|_{\xi = 0} \, .
\end{equation}
By symmetrically ordered, we mean the average of all possible orderings, for example:
\begin{eqnarray}
\label{EqB5}
S\langle \hat{a}^{\dagger}\hat{a} \rangle &=& \frac{1}{2}\left(\hat{a}^\dagger\hat{a} + \hat{a}\hat{a}^\dagger\right) \nonumber \\
S\langle \hat{a}^{\dagger 2}\hat{a}^2 \rangle &=& \frac{1}{6}\left(\hat{a}^{\dagger 2}\hat{a}^2 + \hat{a}^\dagger\hat{a}\hat{a}^\dagger\hat{a}
+ \hat{a}^\dagger\hat{a}^2\hat{a}^\dagger + \hat{a}\hat{a}^{\dagger 2}\hat{a} + \hat{a}\hat{a}^\dagger\hat{a}\hat{a}^\dagger +
\hat{a}^2\hat{a}^{\dagger 2}\right)  \, .
\end{eqnarray}

\section{Thermofields}
\label{thermofields}

It is simplest to consider first an isolated discrete oscillator with annihilation and creation operators $\hat{b}$ and $\hat{b}^\dagger$.  For such an oscillator in a thermal state at temperature $T$ the density operator has the simple, diagonal form:
\begin{equation}
\label{EqC1}
\hat{\rho}_T = (1 - e^{-\beta\hbar\omega})\sum_{n=0}^\infty e^{-n\beta\hbar\omega}|n\rangle\langle n| \, ,
\end{equation}
where $\beta = (k_BT)^{-1}$ is the inverse temperature.  The mean number of excitations is
\begin{equation}
\label{EqC2}
\bar{n} = \frac{1}{e^{\beta\hbar\omega} - 1}
\end{equation}
and we can write the density operator in terms of this mean:
\begin{equation}
\hat{\rho}_T = \frac{1}{\bar{n} + 1}\sum_{n=0}^\infty \left(\frac{\bar{n}}{\bar{n} + 1}\right)^n |n\rangle\langle n| \, .
\end{equation}

The thermofield technique \cite{Takahashi,Umezawa1982,Knight1985,Dalton1987,Umezawa1993,Dalton2015} starts with the observation that we can write a \emph{pure} state that has the same statistical properties as the thermal mixed state (\ref{EqC1}).  To construct this state we consider a doubled state space in which we introduce a second oscillator with annihilation and creation operators $\hat{\tilde{b}}$ and $\hat{\tilde{b}}^\dagger$.  The two-mode pure state, the \emph{thermal vacuum}:
\begin{equation}
\label{EqC4}
|0(\beta)\rangle = (1 - e^{-\beta\hbar\omega})^{1/2}\sum_{n=0}^\infty e^{-\beta\hbar\omega/2}|n,\tilde{n}\rangle
\end{equation}
has precisely the same single-mode properties as the single-mode thermal state:
\begin{equation}
\label{EqC5}
\langle 0(\beta)| f(\hat{b},\hat{b}^\dagger)|0(\beta)\rangle = {\rm Tr}\left(f(\hat{b},\hat{b}^\dagger)\right) \, .
\end{equation}
It is straightforward to show that a similar procedure can be applied can be applied to express any mixed state in terms of a pure state in a doubled state space \cite{Dalton1987}.  When this procedure was rediscovered in quantum information theory, it acquired the name purification \cite{QIbook}.

The benefit of introducing the thermal vacuum state comes from the fact that it is related to the two-mode vacuum state, $|0,\tilde{0}\rangle$, via a unitary transformation:
\begin{eqnarray}
\label{EqC6}
|0(\beta)\rangle &=& \hat{S}(\theta)|0,\tilde{0}\rangle \nonumber \\
&=& \exp\left[\theta(\beta)\left(\hat{\tilde{b}}^\dagger\hat{b}^\dagger - \hat{b}\hat{\tilde{b}}\right)\right]|0,\tilde{0}\rangle \, .
\end{eqnarray}
This transformation produces the desired state if we select $\theta(\beta)$ such that 
\begin{equation}
\label{EqC7}  {\rm sinh}^2\theta(\beta) = \bar{n} \, .
\end{equation}
Readers with a background in quantum optics may recognise $|0(\beta)\rangle$ as a two-mode squeezed vacuum state \cite{Radmore1997,Milburn1984}. The unitary nature of this transformation means that we can convert, by means of the inverse transformation, our effective thermal state into a vacuum state, accompanied by a modified Hamiltonian.  Before we can do this, however, we require a Hamiltonian for the tilde oscillator.  The natural way to introduce this is as an inverted oscillator, so that our free oscillator Hamiltonian becomes
\begin{equation}
\label{EqC8}
\hat{H}_0 = \hbar\Omega_0\left(\hat{b}^\dagger\hat{b} - \hat{\tilde{b}}^\dagger\hat{\tilde{b}}\right) \, ,
\end{equation}
which has the advantage that it commutes with the unitary transformation.  It is essential, moreover, to avoid introducing undesired couplings between the original and the tilde operators.

If our oscillator is coupled to another quantum system via its annihilation and creation operators, then the required unitary transformation effects the replacement
\begin{eqnarray}
\label{EqC9}
\hat{b} &\rightarrow& \hat{S}(\theta)\hat{b}\hat{S}^\dagger(\theta) = \hat{b}\:{\rm cosh}\theta(\beta) - \hat{\tilde{b}}^\dagger {\rm sinh}\theta(\beta)
\nonumber \\
\hat{b}^\dagger &\rightarrow& \hat{S}(\theta)\hat{b}^\dagger\hat{S}^\dagger(\theta) = \hat{b}^\dagger{\rm cosh}\theta(\beta) - \hat{\tilde{b}} \: {\rm sinh}\theta(\beta)
\, .
\end{eqnarray}
As a simple illustration we note that the expectation value of $\hat{b}^\dagger\hat{b}$ is
\begin{eqnarray}
\label{EqC10}
\langle 0(\beta)|\hat{b}^\dagger\hat{b}|0(\beta)\rangle\rangle 
&=& \langle 0,\tilde{0}|(\hat{b}^\dagger {\rm cosh}\theta(\beta) - \hat{\tilde{b}}\:{\rm sinh}\theta(\beta)) \nonumber \\
& & \qquad \times (\hat{b}\: {\rm cosh}\theta(\beta) - \hat{\tilde{b}}^\dagger{\rm sinh}\theta(\beta))|0,\tilde{0}\rangle \nonumber \\
&=& {\rm sinh}^2\theta(\beta) \, .
\end{eqnarray}
In place of a coupling to a single harmonic oscillator in a thermal state with inverse temperature $\beta$, the transformed Hamiltonian has a coupling to a regular oscillator in its ground state with a coupling strength increased by ${\rm cosh}\theta(\beta)$ and a coupling to a second \emph{inverted} oscillator in its most highly excited state, with the original coupling multiplied by ${\rm sinh}\theta(\beta)$.  The inverted oscillator can only inject quanta (at least initially) while the regular oscillator can only extract them.

To complete the picture we need only generalise this description to our continuum operators.  We do this by using our continuum thermal vacuum state in the form
\begin{eqnarray}
\label{EqC11}
|0(\beta)\rangle &=& \hat{S}(\theta[\omega])|0,\tilde{0}\rangle \nonumber \\
&=& \exp\left[\int d\omega \: \theta(\beta,\omega)\left(\hat{\tilde{b}}^\dagger(\omega) \hat{b}^\dagger(\omega) 
- \hat{b}(\omega)\hat{\tilde{b}}(\omega)\right)\right]|0,\tilde{0}\rangle \, ,
\end{eqnarray}
where $|0,\tilde{0}\rangle$ now denotes the doubled continuum vacuum state and $\hat{\tilde{b}}(\omega)$ is the annihilation operator corresponding to \emph{adding} a quantum of frequency $\omega$ to the inverted, tilde reservoir.  As with the discrete oscillator, we can transform into an equivalent vacuum picture with the free Hamiltonian for the reservoir changed to
\begin{equation}
\label{EqC12}
\hat{H}_0 = \int d\omega \: \hbar\omega\left(\hat{b}^\dagger(\omega)\hat{b}(\omega) - 
\hat{\tilde{b}}^\dagger(\omega)\hat{\tilde{b}}(\omega)\right) \, ,
\end{equation}
and the continuum annihilation and creation operators transformed by the inverse unitary transformation
\begin{eqnarray}
\label{EqC13}
\hat{b}(\omega') &\rightarrow& \hat{S}(\theta[\omega])\hat{b}(\omega')\hat{S}^\dagger(\theta[\omega]) 
= \hat{b}(\omega'){\rm cosh}\theta(\beta, \omega') - \hat{\tilde{b}}^\dagger(\omega') {\rm sinh}\theta(\beta,\omega')
\nonumber \\
\hat{b}^\dagger(\omega') &\rightarrow& \hat{S}(\theta[\omega])\hat{b}^\dagger(\omega')\hat{S}^\dagger(\theta[\omega]) 
= \hat{b}^\dagger(\omega'){\rm cosh}\theta(\beta, \omega') - \hat{\tilde{b}}(\omega') {\rm sinh}\theta(\beta,\omega') \, , \nonumber \\
& & 
\end{eqnarray}
so that the expectation value of $\hat{b}^\dagger(\omega)\hat{b}(\omega')$ is 
\begin{eqnarray}
\label{EqC14}
\langle 0(\beta)|\hat{b}^\dagger(\omega)\hat{b}(\omega')|0(\beta)\rangle &=&
\langle 0,\tilde{0}|(\hat{b}^\dagger(\omega) {\rm cosh}\theta(\beta,\omega) - \hat{\tilde{b}}(\omega){\rm sinh}\theta(\beta,\omega)) \nonumber \\
& & \qquad \times (\hat{b}(\omega') {\rm cosh}\theta(\beta,\omega') - \hat{\tilde{b}}^\dagger(\omega'){\rm sinh}\theta(\beta,\omega'))|0,\tilde{0}\rangle \nonumber \\
&=& \bar{n}(\omega)\delta(\omega - \omega') \, .
\end{eqnarray}



\section*{References}
\begin{thebibliography}{10}

\bibitem{Chan2011}
Chan J, Alegre T P M, Safavi-Naeini A H, Hill J T, Gr\"{o}blacher S, Aspelmeyer M and Painter O 2011 Laser Cooling of a Nanomechanical Oscillator into its Quantum Ground State \emph{Nature} {\bf 478} 89--92

\bibitem{Gigan2006}
Gigan S, B\"{o}hm H R, Paternostro M, Blaser F, Langer G, Hertzberg J B, Schwab K C, B\"{a}uerle D, Aspelmeyer M and Zeilinger A 2006 Self-Cooling of a Micromirror by Radiation Pressure \emph{Nature} {\bf 444} 67--70

\bibitem{Stannigel2010}
Stannigel K, Rabl P, S{\o}rensen AS, Zoller P and Lukin M D 2010 Optomechanical Transducers for Long-Distance Quantum Communication \emph{Phys. Rev. Lett.} {\bf 105} 220501

\bibitem{Galve2010}
Galve F, Pach\'{o}n L A and Zueco P 2010 Bringing Entanglement to the High Temperature Limit \emph{Phys. Rev. Lett.} {\bf 105} 180501

\bibitem{Zhang2014}
Zhang K, Bariani F and Meystre P 2014 Quantum Optomechanical Heat Engine \emph{Phys. Rev. Lett.}  {\bf 112} 150602

\bibitem{Joshi}
Joshi C, \"{O}hberg P, Cresser J D and Andersson E 2014 Markovian Evolution of Strongly Coupled Harmonic Oscillators \emph{Phys. Rev.} A {\bf 90} 063815

\bibitem{Brunelli2015}
Brunelli M, Xuereb A, Ferraro A, De Chiara G, Kiesel N and Paternostro M 2015 Out-of-Equilibrium Thermodynamics of Quantum Optomechanical Systems \emph{New J. Phys.} {\bf 17} 035016

\bibitem{Campisi}
Campisi M, Talkner P and H\"{a}nggi P 2009 Fluctuation Theorem for Arbitrary Open Quantum Systems \emph{Phys. Rev. Lett.} {\bf 102} 210401

\bibitem{Talkner}
Talkner P, Campisi M and H\"{a}nggi P 2009 Fluctuation Theorems in Driven Open Quantum Systems \emph{J. Stat. Mech.} P02025

\bibitem{Gemmer}
Gemmer J, Michel M and Mahler G 2004 \emph{Quantum Thermodynamics} (Berlin: Springer)

\bibitem{Vinjanampathy2016}
Vinjanampathy S and Anders J 2016 Quantum Thermodynamics \emph{Contemp. Phys.} {\bf 57} 545--579

\bibitem{Allahverdyan}
Allahverdyan AE and Nieuwenhuizen TM 2000 Extraction of Work from a Single Thermal Bath in the Quantum Regime \emph{Phys. Rev. Lett.} {\bf 85} 1799--1802

\bibitem{CalderiaI}
Caldeira A O and Leggett A J 1983 Path Integral Approach to Quantum Brownian Motion \emph{Physica} {\bf 121 A} 587--616

\bibitem{CaldeiraII}
Caldeira A O and Leggett A J 1983  Quantum Tunnelling in a Dissipative System \emph{Ann. Phys. N.Y.} {\bf 149} 374--456

\bibitem{Senitzky}
Senitzky I R 1960 Dissipation in Quantum Mechanics. The Harmonic Oscillator \emph{Phys. Rev.} {\bf 119} 670--679

\bibitem{Louisell1964}
Louisell W H 1964 \emph{Radiation and Noise in Quantum Electronics} (New York: McGraw-Hill)

\bibitem{Radmore1997}
Barnett S M and Radmore P M 1997 \emph{Methods in Theoretical Quantum Optics} (Oxford: Oxford University Press)

\bibitem{Louisell1973}
Louisell W H 1973 \emph{Quantum Statistical Properties of Radiation} (New York: Wiley)

\bibitem{Grabert1982}
Grabert H 1982 \emph{Projection Operator Techniques in Nonequilibrium Statistical Mechanics} (Berlin: Springer-Verlag)

\bibitem{Perina}
Pe\v{r}ina J 1991 \emph{Quantum Statistics of Linear and Nonlinear Optical Phenomena} 2nd ed. (Dordrecht: Kluwer)

\bibitem{Meystre1991}
Meystre P and Sargent III M 1991 \emph{Elements of Quantum Optics} 2nd ed. (Berlin: Springer-Verlag)

\bibitem{Carmichael1993}
Carmichael H 1993 \emph{An Open Systems Approach to Quantum Optics} (Berlin: Springer-Verlag)

\bibitem{Walls1994}
Walls D F and Milburn G J 1994 \emph{Quantum Optics} (Berlin: Springer-Verlag)

\bibitem{Carmichael1999}
Carmichael H J 1999 \emph{Statistical Methods in Quantum Optics 1} (Berlin: Springer)

\bibitem{Breuer2002}
Breuer H-P and Petruccione F 2002 \emph{The Theory of Open Quantum Systems} (Oxford; Oxford University Press)

\bibitem{Gardiner2004}
Gardiner C W and Zoller P 2004 \emph{Quantum Noise} 2nd ed. (Berlin: Springer)

\bibitem{Ficek2004}
Ficek Z and Swain S 2004 \emph{Quantum Interference and Coherence} (New York: Springer)

\bibitem{Wiseman2010}
Wiseman H M and Milburn G J 2010 \emph{Quantum Measurement and Control} (Cambridge: Cambridge University Press)

\bibitem{Weiss2012}
Weiss U 2012 \emph{Quantum Dissipative Systems} 4th ed. (New Jersey: World Scientific)

\bibitem{Igor1}
Navotn\'{y} J, Alber G and Jex I 2012 Asymptotic Properties of Quantum Markov Chains \emph{J. Phys. A: Math. Theo.} {\bf 45} 485301

\bibitem{Igor2}
Navotn\'{y} J, Mary\v{s}ka J and Jex I 2018 Quantum Markov Processes: from Attractors to Explicit Forms of Asymptotic States \emph{Eur. Phys. J. Plus} {\bf 133} 310

\bibitem{Stenholm2001}
Barnett S M and Stenholm S 2001 Hazards of Reservoir Memory \emph{Phys. Rev.} A {\bf 64} 033808

\bibitem{Maniscalco2007}
Maniscalco S 2007 Complete Positivity of a Spin-1/2 Master Equation with Memory \emph{Phys. Rev.} A {\bf 75} 062103

\bibitem{Haikka2011}
Kaikka P, Cresser J D and Maniscalco S 2011 Comparing Different Non-Markovianity Measures in a Driven Qubit System \emph{Phys. Rev.} A {\bf 83} 012112

\bibitem{Breuer2012}
Breuer H-P 2012 Foundations and Measures of Non-Markovianity \emph{J. Phys. B: At. Mol. Opt. Phys.} {\bf 45} 154001

\bibitem{Pernice2012}
Pernice A, Helm J and Strunz W T 2012 System-Environment Correlations and Non-Markovian Dynamics \emph{J. Phys. B: At. Mol. Opt. Phys.} {\bf 45} 154007

\bibitem{Vacchini2012}
Vacchini  B 2012 A Classical Appraisal of Quantum Definitions of Non-Markovian Dynamics \emph{J. Phys. B: At. Mol. Opt. Phys.} {\bf 45} 154007

\bibitem{Chrascinski2014}
Chra\'{s}in\'{n}ski D and Maniscalco S 2014 Degree of Non-Markovianity of Quantum Evolution \emph{Phys. Rev. Lett.} {\bf 112} 120404

\bibitem{Hall2014}
Hall M J W, Cresser J D, Li L and Andersson E 2014 Canonical Form of Master Equations and Characterization of Non-Markovianity \emph{Phys. Rev.} A {\bf 89} 042120

\bibitem{Hu}
Hu B L, Paz J P and Zhang Y 1992 Quantum Brownian Motion in a General Environment: Exact Master Equation with Nonlocal Dissipation and Colored Noise \emph{Phys. Rev.} D {\bf 45} 2843--2861

\bibitem{Diosi1993}
Diosi L 1993 Caldeira-Leggett Master Equation and Medium Temperatures \emph{Physica} {\bf 199} 517--526

\bibitem{Grabert1984}
Grabert H and Weiss U 1984 Quantum Theory of the Damped Harmonic Oscillator \emph{Z. Phys. B - Cond. Matt.} {\bf 55} 87--94

\bibitem{Ford1985}
Ford G W, Lewis J T and O'Connell R F 1985 Quantum Oscillator in a Blackbody Radiation Field \emph{Phys. Rev. Lett.} {\bf 55} 2273--2276

\bibitem{Ford1985a}
Ford G W, Lewis J T and O'Connell R F 1985 Quantum Langevin Equation \emph{Phys. Rev.} A {\bf 37} 4419--4428

\bibitem{Feynman}
Feynman R and Hibbs A R 2010 \emph{Quantum Mechanics and Path Integrals} (Emended ed.) (New York: Dover)

\bibitem{Zinn-Justin}
Zinn-Justin J 2002 \emph{Quantum Field Theory and Critical Phenomena} (Oxford: Oxford University Press)

\bibitem{Kleinert}
Kleinert H 2009 \emph{Path Integrals in Quantum Mechanics, Statistics, Polymer physics and Financial markets} (fifth ed.) (New Jersey: World Scientific)

\bibitem{Smith1987}
Smith C M and Caldeira A O 1987 Generalized Feynman-Vernon Approach to Dissipative Quantum Systems
\emph{Phys. Rev.} A {\bf 36} 3509--3511

\bibitem{Grabert1988}
Grabert H, Schramm P and Ingold G-L 1988 Quantum Brownian Motion: the Functional Integral Approach
\emph{Phys. Rep.} {\bf 168} 115--207

\bibitem{Abrikosov}
Abrikosov A A, Gorkov L P and Dzaloshinski I E 1975 \emph{Methods of Quantum Field Theory in Statistical Physics} (New York: Dover)

\bibitem{Lifshitz}
Lifshitz E M and Pitaevskii L P 1980 \emph{Statistical Physics part 2} (Oxford: Butterworth Heinemann)

\bibitem{Le Bellac}
Le Bellac M 1996 \emph{Thermal Field Theory} (Cambridge: Cambridge University Press)

\bibitem{Takahashi}
Takahashi Y and Umezawa H 1975 Thermo Field Dynamics \emph{Collect. Phenom.} {\bf 2}
55--80.  Reprinted in 1996 \emph{Int. J. Theo. Phys.} {\bf 10} 1755--1805

\bibitem{Umezawa1982}
Umezawa H, Matsumoto H and Tachiki M 1982 \emph{Thermo Field Dynamics and Condensed States}
(Amsterdam: North-Holland)

\bibitem{Knight1985}
Barnett S M and Knight P L 1985 Thermofield Analysis of Squeezing and Statistical Mixtures in Quantum
Optics \emph{J. Opt. Soc. Am.} B {\bf 2} 467--479

\bibitem{Dalton1987}
Barnett S M and Dalton B J 1987 Liouville Space Description of Thermofields and their Generalisations
\emph{J. Phys. A: Math. Gen.} {\bf 20} 411--418

\bibitem{Umezawa1993}
Umezawa H 1993 \emph{Advanced Field Theory, Micro, Macro and Thermal Physics} (New York: American 
Institute of Physics)

\bibitem{Dalton2015}
Dalton B J, Jeffers J and Barnett S M 2015 \emph{Phase Space Methods for Degenerate Quantum Gases}
(Oxford: Oxford University Press)

\bibitem{Subasi2012}
Suba\c{s}{\i} Y, Fleming C H , Taylor J M, and Hu B L 2012 Equilibrium states of open quantum systems in the strong coupling regime \emph{Phys. Rev.} E {\bf 86} 061132

\bibitem{Haake}
Haake F and Reibold R 1985 Strong Damping and Low-Temperature Anomalies for the Harmonic Oscillator \emph{Phys. Rev.} A {\bf 32} 2462--2475

\bibitem{O'Connell}
Ford G W, Lewis J T and O'Connell R F 1988 Independent Oscillator Model of a Heat Bath: Exact Diagonalization of the Hamiltonian \emph{J. Stat. Phys.} {\bf 53} 439--455

\bibitem{preprint}
Barnett S M, Cresser J D and Croke S 2015 Theory of the Strongly-Damped Quantum Harmonic Oscillator 
\emph{arXiv}1508:02442 [quant-ph]

\bibitem{Huttner1992a}
Huttner B and Barnett S M 1992 Dispersion and Loss in a Hopfield Dielectric \emph{Europhys. Lett.} {\bf 18} 487--492

\bibitem{Huttner1992b}
Huttner B and Barnett S M 1992 Quantization of the Electromagnetic Field in Dielectrics \emph{Phys. Rev.} A {\bf 46} 4306--4322

\bibitem{Dutra}
Dutra S M 2005 {\it Cavity Quantum Electrodynamics} (Hoboken New Jersey: Wiley)

\bibitem{PhilbinAnders}
Philbin T G and Anders J 2016 Thermal Energies and Quantum Damped Oscillators Coupled to Reservoirs
\emph{J. Phys. A: Math. Theo.} {\bf 49} 215303

\bibitem{Philbin2012}
Philbin T G 2012 Quantum Dynamics of the Damped Harmonic Oscillator \emph{New J. Phys.} {\bf 14} 083043

\bibitem{Schrodinger}
Schr\"{o}dinger E 1926 Quantisierung als Eigenwertproblem \emph{Ann. Phys.} {\bf 79}, 489-527; reprinted in translation as ``Quantisation as a Problem of Proper Values (Part II)" in Schr\"{o}dinger E 1982 \emph{Collected Papers on Wave Mechanics} 3rd ed. (Providence, Rhode Island: AMS Chelsea Publishing)

\bibitem{Langevin}
Langevin P 1908 Sur la Th\'{e}orie du Mouvement Brownien \emph{Comptes Rend. Acad. Sci. (Paris)} {\bf 146} 530--533; reprinted in a translation by Gythiel A 1997 ``On the Theory of Brownian Motion" \emph{Am. J. Phys.} {\bf 65} 1079--1081

\bibitem{Uhlenbeck1930}
Uhlenbeck G E and Ornstein L S 1930 On the Theory of Brownian motion \emph{Phys. Rev.} {\bf 36}, 823--841

\bibitem{Wang1945}
Wang M C and Uhlenbeck G E 1945 On the Theory of Brownian motion II \emph{Rev. Mod. Phys.} {\bf 17} 323--342

\bibitem{Risken}
Risken H 1989 \emph{The Fokker-Planck Equation} 2nd ed. (Berlin: Springer)

\bibitem{Lemons}
Lemons D S 2002 \emph{An Introduction to Stochastic Processes in Physics} (Baltimore: Johns Hopkins University Press)

\bibitem{Mazo}
Mazo R M 2002 \emph{Brownian Motion} (Oxford: Oxford University Press)

\bibitem{Agarwal}
Agarwal G S 1971 Brownian Motion of a Quantum Oscillator \emph{Phys. Rev.} A {\bf 4} 739--747

\bibitem{UllersmaI}
Ullersma P 1966 An Exactly Solvable Model for Brownian Motion I. Derivation of the Langevin Equation \emph{Physica} {\bf 32} 27-55

\bibitem{UllersmaII}
Ullersma P 1966 An Exactly Solvable Model for Brownian Motion II. Derivation of the Fokker-Planck equation and the Master Equation \emph{Physica} {\bf 32} 56--73

\bibitem{Kheirandish2006}
Kheirandish F and Amooshahi 2006 Electromagnetic Field Quantisation in a Linear Polarizable and Magnetizable Medium \emph{Phys. Rev.} A {\bf 74} 042102

\bibitem{Kheirandish2008}
Kheirandish F and Soltani M 2008 Extension of the Huttner-Barnett Model to a Magnetodielectric Medium
\emph{Phys. Rev.} A {\bf 78} 012102 

\bibitem{Kheirandish2009}
Kheirandish F, Amooshahi M and Soltani M 2009 A Canonical Approach to Electromagnetic Field Quantization
in a Nonhomogeneous and Anistropic Magnetodielectric Medium \emph{J. Phys. B: At. Mol. Opt.} {\bf 42}
075504 

\bibitem{Amooshahi2009}
Amooshahi M 2009 Canonical Quantisation of the Electromagnetic Field in an Anisotropic Polarizable and Magnetizable Medium \emph{J. Math. Phys.} {\bf 50} 062301 

\bibitem{Munro}
Munro W J and Gardiner C W 1996 Non-Rotating-Wave Master Equation \emph{Phys. Rev.} A {\bf 53} 2633--2640

\bibitem{Stenholm}
Stenholm S 1996 Simple Quantum Dynamics in \emph{Quantum Dynamics of Simple Systems} G. L. Oppo, S. M. Barnett, E. Riis and M. Wilkinson eds. (Bristol: Scottish Universities Summer Schools in Physics and Institute of Physics) pp. 267--313

\bibitem{Cresser2005}
Barnett S M and Cresser J D 2005 Quantum Theory of Friction \emph{Phys. Rev.} A {\bf 72} 022107 

\bibitem{Cresser2006}
Barnett S M, Jeffers J and Cresser J D 2006 From Measurements to Quantum Friction \emph{J. Phys: Condens. Matter} {\bf 18} S401--S410

\bibitem{Bixon}
Bixon M and Jortner J 1968 Intramolecular Radiationless Transitions \emph{J. Chem. Phys.} {\bf 48} 715--725

\bibitem{Kyrola}
Kyr\"{o}la E 1984 Photoexcitation of a Quasi-Continuum: Connections to Few-Level Dynamics \emph{J. Opt. Soc. Am.} B {\bf 1} 737--74

\bibitem{Milonni}
Milonni P W, Ackerhalt J R, Galbraith H W and Shil M-L 1988 Exponential Decay and Quantum Mechanical Spreading in a Quasicontinuum Model \emph{Phys. Rev.} A {\bf 28} 32--39

\bibitem{Eberly}
Eberly J H, Yeh J J and Bowden C M 1982 Interupted Coarse-Grained Theory of Quasi-Continuum Photoexcitation \emph{Chem. Phys. Lett.} {\bf 86} 76--80

\bibitem{Tarzi}
Radmore P M and Tarzi S 1987 Photo-Excitation of a Structured Continuum: a Soluble Model \emph{J. Mod. Opt.} {\bf 34} 1409--1420

\bibitem{Bohr} 
Bohr H 2018 \emph{Almost Periodic Functions} (New York: Dover)

\bibitem{Fano35}
Fano U 1935 Sullo Spettro di Assortbimento dei Gas Nobili Presso il Limite dello Spettro d'Arco
\emph{Il Nuovo Cimento} {\bf 12} 154--161

\bibitem{Radmore1988}
Barnett SM and Radmore PM 1988 Quantum Theory of Cavity Quasimodes \emph{Opt. Commun.} {\bf 68} 364--368

\bibitem{Georgievskii}
Georgievskii Y and Pollak E 1995 Activated Rate Processes: Anharmonic Corrections to the Quantum Rate \emph{J. Chem. Phys.} {\bf 103} 8910--8920

\bibitem{Ratchov}
Ratchov A, Faure F and Hekking F WJ 2005 Loss of Quantum Coherence in a System Coupled to a Zero-Temperature Environment \emph{Eur. J. Phys.} B  {\bf 46} 519--528

\bibitem{Ghosh}
Ghosh A, Bandyopadhyay M, Dattagupta S and Gupta S 2023 Quantum Brownian Motion: A Review 
arXiv:2306.02665

\bibitem{Franco}
Compagno C, Passante R and Persico F 1995 \emph{Atom-Field Interactions and Dressed Atoms} (Cambridge:
Cambridge University Press)

\bibitem{Casimir}
Casimir H B G and Polder D 1948 The Influence of Retardation on the London-van der Waals Forces 
\emph{Phys. Rev.} {\bf 73} 360--372

\bibitem{Edwin} 
Power E A 1964 \emph{Introductory Quantum Electrodynamics} (London: Longmans)

\bibitem{Craig}
Craig D P and Thirunamachandran T 1984 \emph{Molecular Quantum Electrodynamics} (London: Academic Press)

\bibitem{Rodney2006}
Loudon R and Barnett S M 2006 Theory of the Linear Polarizability of a Two-Level Atom \emph{J. Phys. B: At. Mol. Opt. Phys.} {\bf 39} S555--S563

\bibitem{Milonni2008}
Milonni PW, Loudon R, Berman PR and Barnett SM 2008 Linear Polarizabilities of Two- and Three- Level Atoms \emph{Phys. Rev.} A {\bf 77} 043835

\bibitem{Philbin2013}
Philbin T G and Horsley S A R 2013 Damping the Zero-Point Energy of a Harmonic Oscillator \emph{arXiv}:
1304.0977v2 [quant-ph]

\bibitem{Araki}
Araki H and Lieb E H 1970 Entropy iIequalities \emph{Commun. Math. Phys.} {\bf 18} 160--170

\bibitem{Wehrl}
Wehrl A 1978 General Properties of Entropy \emph{Rev. Mod. Phys.} {\bf 50} 221--260

\bibitem{QIbook}
Barnett S M 2009 \emph{Quantum Information} (Oxford: Oxford University Press)

\bibitem{Simon}
Barnett S M and Phoenix S J D 1989 Entropy as a Measure of Quantum Optical Correlation \emph{Phys. Rev.} A {\bf 40} 2404--2409

\bibitem{Kirkwood}
Kirkwood J G 1935, Statistical mechanics of fluid mixtures, J. Chem. Phys. {\bf 3}, 300

\bibitem{Trushechkin}
Trushechkin A S, Merkli M, Cresser J D, and Anders J 2022 AVS Quantum Science {\bf 4}, 012301

\bibitem{Martinez} 
Martinez E A and Paz J P 2013 Dynamics and Thermodynamics of Linear Quantum Open Systems \emph{Phys. Rev. Lett.} {\bf 110} 130406

\bibitem{Joan2011}
Vaccaro J A and Barnett S M 2011 Information Erasure without an Energy Cost \emph{Proc. R. Soc.} A {\bf 467} 1770--1778

\bibitem{Joan2013}
Barnett S M and Vaccaro J A 2013 Beyond Landauer Erasure \emph{Entropy} {\bf 15} 4956--4968

\bibitem{Roncaglia}
Roncaglia A J, Cerisola F and Paz J P 2014 Work as a Generalized Measurement \emph{Phys. Rev. Lett.}
{\bf 113} 0250601 

\bibitem{Binder}
Binder F, Vinjanampathy S, Modi K and Goold J 2015 Operational Thermodynamics of Open Quantum Systems 
\emph{Phys. Rev.} E {\bf 91} 032119

\bibitem{Brandao}
Brandao F, Horodecki M, Ng K, Oppenheim J and Wehner S 2015 The Second Laws of Quantum Thermodynamics
\emph{Proc. Natl. Acad. Sci. USA} {\bf 112} 3275--3279

\bibitem{Fano}
Fano U 1956 Effects of Configuration Interaction on Intensities and Phase Shifts \emph{Phys. Rev.} {\bf 124} 1866--1878

\bibitem{Dirac1927}
Dirac P A M 1927 \"{U}ber die Quantenmechanik der Sto{\ss}vorg\"{a}nge \emph{Z. Phys.} {\bf 44} 585--595; reprinted in translation as ``On the Quantum Mechanics of Collisions" in R. H. Dalitz (ed.) 1995 \emph{The Collected Works of P. A. M. Dirac 1924--1948} (Cambridge: Cambridge University Press)

\bibitem{Cahill}
Cahill K E and Glauber R J 1969 Ordered Expansions of Boson Amplitude Operators \emph{Phys. Rev.}
{\bf 177} 1857--1881

\bibitem{Dalton2023}
Barnett S M and Dalton B J 2023 Glauber-Sudarshan P-Representations for Fermions \emph{Phys. Scr.}
{\bf 98} 044006

\bibitem{Weyl}
Weyl H 1950 \emph{The Theory of Groups and Quantum Mechanics} (New York: Dover) p. 274

\bibitem{Milburn1984}
Milburn G J 1984 Multimode Minimum Uncertainty Squeezed States \emph{J. Phys. A: Math. Gen.} {\bf 17} 737--745


\end{thebibliography}
\end{document}